\documentclass[useAMS,usenatbib]{mn2e}
\usepackage{epsfig}

\def\lesssim{\mathrel{\hbox{\rlap{\hbox{\lower4pt\hbox{$\sim$}}}\hbox{$<$}}}}
\def\gtrsim{\mathrel{\hbox{\rlap{\hbox{\lower4pt\hbox{$\sim$}}}\hbox{$>$}}}}
\def\la{\mathrel{\hbox{\rlap{\hbox{\lower4pt\hbox{$\sim$}}}\hbox{$<$}}}}
\def\ga{\mathrel{\hbox{\rlap{\hbox{\lower4pt\hbox{$\sim$}}}\hbox{$>$}}}}

\def\spose#1{\hbox to 0pt{#1\hss}}
\def\approxlt{\mathrel{\spose{\lower 3pt\hbox{$\sim$}}
	\raise 2.0pt\hbox{$<$}}}
\def\approxgt{\mathrel{\spose{\lower 3pt\hbox{$\sim$}}
	\raise 2.0pt\hbox{$>$}}}

\def\<{\thinspace}
\def\s{\hbox{\phantom{5}}}	
\def\ss{\s\s}		

\def\boxit#1{\vbox{\hrule\hbox{\vrule\kern3pt\vbox{\kern3pt
          #1 \kern3pt}\kern3pt\vrule}\hrule}}

\def\s{{\rm\thinspace s}}

\def\h50{\hbox{$\rm\thinspace h_{50}$}}
\def\h50m1{\hbox{$\rm\thinspace h_{50}^{-1}$}}

\title[The final PAndAS catalogue of M31 outer halo GCs]{The outer halo globular cluster system of M31 -- I. The final PAndAS catalogue }
\author[Huxor et al.]{A. P. Huxor$^{1}$, A. D. Mackey$^{2}$, A. M. N. Ferguson$^{3}$, M. J. Irwin$^{4}$, N. F. Martin$^{5}$, 
\newauthor N. R. Tanvir$^{6}$, J. Veljanoski$^{3}$, A. McConnachie$^{7}$, C. K. Fishlock$^{2}$, R. Ibata$^{5}$,  
\newauthor G. F. Lewis$^{8}$\\
$^{1}$Astronomisches Rechen-Institut, Zentrum f\"{u}r Astronomie der Universit\"{a}t Heidelberg, M\"{o}nchhofstra{\ss}e 12 - 14, \\69120 Heidelberg, Germany.\\
$^{2}$Research School of Astronomy \& Astrophysics, Australian National University, Mt. Stromlo Observatory, Cotter Road,\\ Weston Creek, ACT 2611, Australia\\
$^{3}$SUPA, Institute for Astronomy, University of Edinburgh, Royal Observatory, Blackford Hill, Edinburgh EH9 3HJ, UK\\
$^{4}$Institute of Astronomy, Madingley Road, Cambridge, CB3 0HA \\
$^{5}$Observatoire de Strasbourg, 11, rue de l'Universit\'{e}, F-67000, Strasbourg, France \\
$^{6}$Department of Physics and Astronomy, University of Leicester, University Road, Leicester LE1 7RH\\
$^{7}$NRC Herzberg Institute of Astrophysics, 5071 West Saanich Road, Victoria, British Columbia V9E 2E7, Canada\\
$^{8}$Institute of Astronomy, School of Physics, A29, University of Sydney, NSW 2006, Australia}

\begin{document}

\date{}

\pagerange{\pageref{firstpage}--\pageref{lastpage}} \pubyear{2014}

\maketitle

\label{firstpage}

\begin{abstract}
We report the discovery of 59 globular clusters (GCs) and two candidate GCs in a search of the halo of M31, primarily via visual inspection of CHFT/MegaCam imagery from the Pan-Andromeda Archaeological Survey (PAndAS). The superior quality of these data also allow us to check the classification of remote objects in the Revised Bologna Catalogue (RBC), plus a subset of GC candidates drawn from SDSS imaging. We identify three additional new GCs from the RBC, and confirm the GC nature of 11 SDSS objects (8 of which appear independently in our remote halo catalogue); the remaining $188$ candidates across both lists are either foreground stars or background galaxies. Our new catalogue represents the first uniform census of GCs across the M31 halo -- we find clusters to the limit of the PAndAS survey area at projected radii of up to $R_{\rm proj} \sim 150$ kpc. Tests using artificial clusters reveal that detection incompleteness cuts in at luminosities below $M_V = -6.0$; our 50\% completeness limit is $M_V \approx -4.1$. We construct a uniform set of PAndAS photometric measurements for all known GCs outside $R_{\rm proj} = 25$ kpc, and any new GCs within this radius. With these data we update results from \citet{Huxoretal11}, investigating the luminosity function (LF), colours and effective radii of M31 GCs with a particular focus on the remote halo. We find that the GCLF is clearly bimodal in the outer halo ($R_{\rm proj} > 30$ kpc), with the secondary peak at $M_V \sim -5.5$. We argue that the GCs in this peak have most likely been accreted along with their host dwarf galaxies. Notwithstanding, we also find, as in previous surveys, a substantial number of GCs with above-average luminosity in the outer M31 halo -- a population with no clear counterpart in the Milky Way.
\end{abstract}


\section{Introduction}
Globular cluster (GC) systems are thought to trace both major star-formation episodes and accretion events. As such they have proven to be valuable tools for the study of their host galaxies \citep{Georgievetal12} -- from the seminal Milky Way (MW) work of \citet{SearleZinn78} to recent studies of more distant galaxies \citep{forteetal12,forbesetal11,ChiesSantosetal11}.  

The GC system of M31 has naturally been the focus of particular interest, providing  (as a massive spiral galaxy) an excellent comparison to our own Milky Way. Moreover, the proximity of M31 (at $\sim 780$ kpc)\footnote{Throughout this paper we use the distance to M31 from \citet{McConnachieetal05}; see also \citet{Connetal12}.} allows for detailed investigation of its GC populations, which have been extensively studied  (e.g. \citealt{Cramptonetal85,Battistinietal87,Elson88,Huchraetal91,Barmbyetal00,Perrettetal02,Fanetal08,galleti:09,Caldwelletal09,Fanetal10,Caldwelletal11}). Most of these studies have  dealt with the regions comparatively close to the centre of M31, typically within $20-25$ kpc in projection. This is because the relative proximity of M31 also poses a problem in that the full extent of its stellar halo subtends a substantial angle on the sky ($\ga 20\degr$ in diameter) which is difficult to search uniformly for GCs, especially those with low luminosities and/or surface brightnesses. The Pan-Andromeda Archaeological Survey \citep[PAndAS;][]{McConnachieetal09} almost completely obviates these issues: its imaging spans a very wide area, typically reaching a projected distance $R_{\rm proj} \sim 150$ kpc from M31\footnote{Although this is still some distance short of the likely virial radius of M31.} -- and is yet sufficiently deep to allow the identification of even faint GCs.

With high quality wide-field imaging such as that obtained for PAndAS, M31 halo GCs are much more easily located than those in more central regions where the background and crowding due to the M31 disk hinders reliable identification of star clusters in ground-based data. Halo GCs also offer the opportunity to study regions with very long dynamical time-scales that can preserve evidence of past events. If formed {\it in-situ}, remote halo GCs will have been much less affected by tidal forces than those towards the centre; if accreted along with dwarf satellite galaxies, their properties may reflect the nature of the original hosts.

This paper continues and extends earlier investigations of the GC population of M31 by our group. In particular, it provides the final catalogue of halo GCs from PAndAS, greatly extending our previous surveys and results -- specifically those of \citet{Huxoretal08} (hereafter,  Hux08) and \citet{Huxoretal11} (hereafter,  Hux11).  In Hux08 we presented 40 new GCs from a precursor survey to PAndAS conducted using the Wide-Field Camera (WFC) on the Isaac Newton Telescope (INT) along with some early imaging from MegaCam on the Canada-France-Hawaii Telescope (CFHT), and updated the classifications of many entries in the Revised Bologna Catalog (RBC)\footnote{http://www.bo.astro.it/M31/} --  the most complete catalogue of M31 GCs, and widely used by the community\footnote{Note that at that time we worked with Version 3.0 of the RBC; for the present work we refer to Version 5 from August 2012}.  Hux11 explored the ``ensemble" properties of the updated M31 GC sample from Hux08. In the present paper we exploit the full, final PAndAS data, searching for new GCs, investigating candidate GCs from the RBC, and updating many of the results from  Hux11 with a particular focus on the properties of the GCs in the halo.

In addition to M31, the PAndAS data (and its preceding INT/WFC survey) also extend to M33, and our work on the GCs in this galaxy is published elsewhere \citep{Huxoretal09, Cockcroftetal11}.  We have also used PAndAS imaging to discover new GCs in the M31 dwarf elliptical (dE) satellites NGC 147 and NGC 185 (three GCs and one GC respectively), as described in \citet{Veljanoskietal13b}. Although, strictly speaking, these clusters reside within the halo of M31, we do not include them in the present paper as they possess clearly identified (and intact) host galaxies.

The GCs listed in our previous catalogue (Hux08)  provided targets for follow-up observations and analysis, both by our own group and by others. In particular, our {\it Hubble Space Telescope} ({\it HST}) observations of many of the halo GCs led to a number of studies of their colour-magnitude diagrams (CMDs) and structural properties \citep{Mackeyetal07,Perinaetal09,Perinaetal11,Tanviretal12,Federicietal12,Perinaetal12,WangMa12}. Many of those GCs were also observed spectroscopically with ground-based facilities -- for example, \citet{AlvesBritoetal09} observed several at high resolution with the Keck Telescope. Similarly,  \citet{Ma12} used optical and 2MASS photometry of many of our GCs to estimate their ages, masses and metallicities. 

The present paper is the first of a series of works in which we use our catalogue to shed new light on the outer regions of the M31 halo. In an accompanying paper \citep{Veljanoskietal14} we investigate the kinematics of the remote GC system, while in two forthcoming works we will explore the relationship between the GCs and the underlying field halo, and the resolved properties of the GCs through {\it HST} imaging (Mackey et al. in prep).

This paper proceeds as follows: in  \S \ref{data_and_search} we describe the CFHT/MegaCam dataset we employed, and the strategy used to locate new GCs. The newly discovered GCs are then presented in \S \ref{new_clusters}. In addition to discovering new GCs, we also used the same imaging data to clean previous samples of published M31 GCs and GC candidates, and the results of this undertaking are given in \S \ref{other_catalogue_updates}. The photometry of our new clusters, and all other GCs with a galactocentric distance of greater than 25 kpc, is described and tabulated in \S \ref{photometry}. Next we assess the completeness of our sample, critical to proper exploitation of the catalogue, in \S \ref{completeness}. Finally, in \S \ref{analysis}, we analyse the ensemble photometric properties of the M31 outer halo GC system, using our enlarged and improved catalogue.

\section{The Globular Cluster Survey}
\label{data_and_search}

\subsection{The Data}
The images and photometric catalogue employed in this study were taken from the now-completed PAndAS survey of M31, conducted using the CFHT on Mauna Kea, Hawaii. Details of this survey and its precursors  can be found in a number of previous works \citep[e.g.,][]{McConnachieetal09,Martinetal06,Ibataetal07,McConnachieetal08,Ibataetal13}, but we briefly summarise the key points here. The PAndAS imaging was undertaken with the MegaPrime/MegaCam camera mounted on the CFHT, which comprises 36 CCDs (each 2048 $\times$ 4612 pixels). Each pointing provides a usable field-of-view of 0.96 $\times$ 0.94 deg$^{2}$. Three dithered $450$s sub-exposures in each of the MegaCam $g$ and $i$ filters typically reach  $g \approx$ 26.0 and $i \approx$ 24.8 (for point sources at the $5\sigma$ detection limit) once reduced and combined. 

Crucial to our identification of GCs is the excellent PAndAS image quality. Many early exposures with relatively poor seeing were re-observed towards the end of the survey program, resulting in a mean seeing of $0\farcs67$ in the $g$-band and $0\farcs60$ in the $i$-band, with an rms scatter between frames of $0\farcs10$ in both cases.

After initial reduction of the data at the CFHT, further image processing, calibration, and photometric measurements were conducted at the Cambridge Astronomical Survey Unit (CASU).  The CASU pipeline created final stacked $g$- and $i$-band images at each pointing, and a merged catalogue providing photometric data and star/galaxy classification for all detected sources, both stellar and non-stellar. The complete contiguous survey footprint, comprising 406 individual pointings, reaches to a projected distance of $\sim150$ kpc from M31 in most directions, thus encompassing almost the entire halo. This region is joined to a smaller area around M33, extending to $\sim 50$ kpc from the centre of that galaxy. 

\subsection{Search Strategy}
\begin{figure*}
 \centering
 \includegraphics[width=130mm]{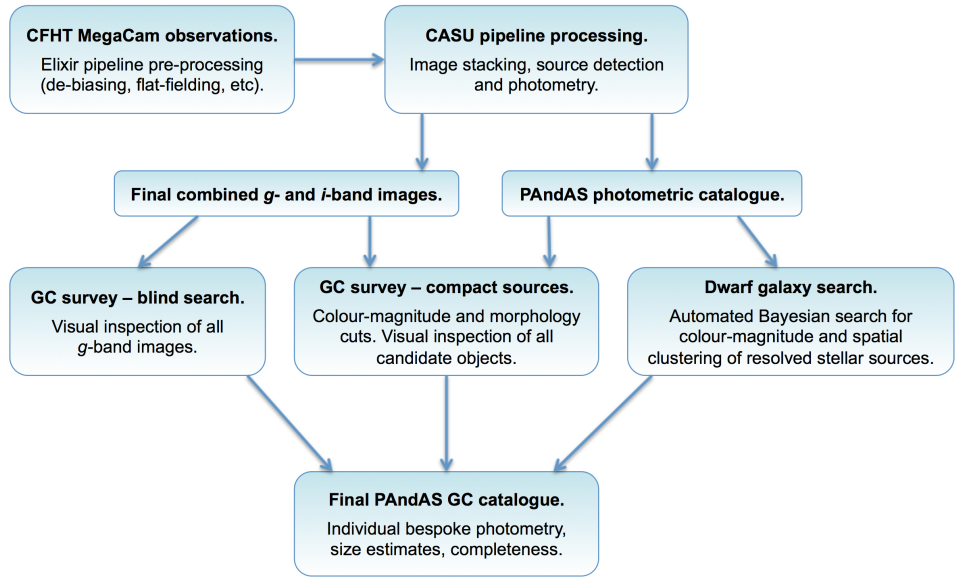}
 \vspace{2pt}
 \caption[schematic of search strategy]
 {A schematic summary of our multi-strand GC search strategy.}\label{Fi:schematic}
\end{figure*}

We adopted a multi-strand search strategy based on our experience from Hux08, in which we found both classical ``compact" M31 GCs, and also the more diffuse ``extended" clusters. Our methodology is summarised in Figure \ref{Fi:schematic}.

GC candidates were selected from the PAndAS photometric catalogue based on their magnitude and colour. Known GCs (both compact and extended) inhabit a broad range of absolute magnitudes and colours ($-10.5 < M_{V} < -3.5$, and $0.0 < (V-I)_{0} < 1.7$) -- limits which we converted to apparent MegaCam $g$ and $i$-magnitudes by using the inverse of the transformation equations (1 to 4) described in Section \ref{photometry}, below, and assuming an M31 distance modulus of $24.47$ and a typical foreground extinction $E(B-V) \sim 0.075$. We further required that any objects selected within these bounds have a non-stellar flag from the CASU photometric pipeline to be considered a GC candidate. This is appropriate for compact M31 GCs, which are always marginally resolved when the image quality is $\sim 0\farcs6 - 0\farcs7$. Diffuse clusters, however, tend not to appear in the catalogue as a single source and can therefore easily be missed with this approach -- we adopted additional search techniques for these objects (see below). Note that the CASU pipeline is not optimised for non-stellar source photometry. Hence, although the magnitudes and colours are sufficiently accurate to identify likely compact GC candidates (especially given our very generous ranges for both), we subsequently undertook our own bespoke photometry of each GC we discovered (see section \ref{photometry}).

We visually inspected a $g$-band image of every candidate object\footnote{Our previous experience revealed that the $g$-band is both more effective and more efficient for identifying GCs than the $i$-band. This is largely due to the greatly reduced prominence of the main contaminants -- background elliptical galaxies and foreground dwarf stars -- in the blue.} and its local surrounding area, using a FITS image viewer to overlay (and so highlight) the positions of the GC candidates with graphic markers. This ensured that adequate attention was drawn to both the less luminous and the more compact candidates. At the distance of M31, and with the high quality of the MegaCam images, GCs generally take the form of a core that is slightly broader than the stellar point-spread function (PSF), surrounded by resolved red giant branch (RGB) stars. This results in an easily distinguished local ``halo" of such objects in well populated-clusters, and/or a broadened core with an irregular appearance for less luminous examples. In almost all cases we found it straightforward to unambiguously classify a GC candidate as a cluster or not. However, the search efficiency was low -- in the vast majority of instances the candidates turned out to be distant background galaxies.

Extended diffuse clusters \citep{Huxoretal05} are problematic because they are typically semi- or completely resolved across their full spatial extent in the MegaCam imaging and thus are not flagged in the PAndAS photometric catalogue by the presence of a single unresolved source. In most cases, however, such objects are also not sufficiently well populated or sufficiently uncrowded to appear as co-located groups of similar stars that could be detected by means of automated algorithms. Our previous experience (Hux08) showed that the most efficient and least biased way to detect such objects is by simple visual inspection of the full survey area. Although labour intensive, this inspection, conducted by APH, led to the discovery of many clusters ($\sim$30\%--40\% of our final sample) that would have otherwise been missed. In addition, it allowed us to independently confirm the nature of the compact clusters previously identified as described above, and ensure that no exceptional such objects lying outside the colour-magnitude selection box were missed. For quality control, to try and minimise the effects of human error, secondary inspection of roughly 30\% of the images was carried out separately by ADM. As a final measure, we looked for cluster detections in the automated search for M31 dwarf spheroidal galaxies conducted by NFM \citep{Martinetal13}. These authors made use of both the spatial and colour-magnitude information of sources, in a probabilistic approach, to identify over-densities of stars with similar photometric properties. We quantitatively assess the completeness of our overall search strategy in Section \ref{completeness}.

\subsection{Spatial Coverage}
Our search covered the PAndAS footprint to its largest radial extent, which ranges from $\sim$120--150 kpc in projection depending on the position angle relative to the M31 centre.  However, the inner extent of our search is less clearly defined. The region which we examined uniformly includes the full area covered by the earlier INT/WFC survey as described in  Hux08. As can be seen from Figure 4 of that paper, the inner region of the INT survey extended to an ellipse with a semi-major axis of 2$^{\circ}$ ($\sim$27 kpc), and an inclination of 77.5$^{\circ}$. Within this region variable crowding makes it difficult to conduct a uniform search for GCs, particularly affecting the discovery of low-luminosity compact GCs, and all extended GCs. Taking the above into account, we expect the completeness limits derived in Section \ref{completeness} to be applicable outside a projected galactocentric radius of $\sim 25$ kpc (although we did locate a handful of GCs within this radius).

\section{The newly discovered clusters}
\label{new_clusters}

Following the procedure described in Section \ref{data_and_search}, we discovered 58 previously unknown GCs and two additional GC candidates\footnote{We note that we subsequently also discovered one additional GC (PA-59), as detailed in Section \ref{RBCupdate}.}. All but one of these came from the independent visual inspection of (i) candidates, and (ii) the full survey area. The exception, PAndAS-31, was discovered via the automated search for M31 dwarf spheroidal galaxies \citep[see][]{Martinetal13}. Note that, as described in that paper, the automated search also uncovered a small subset of the objects discovered independently in our search by eye.

\begin{table}
 \centering
\begin{minipage}{160mm}
 \caption{Locations of newly-discovered PAndAS GCs.}\label{tab:locations}
 \vspace{2pt}
\begin{tabular}{@{}lccr@{}}
\hline \hline
 ID &  \multicolumn{2}{c}{Position (J2000.0)} & $R_{\rm proj}$ \\
   &  RA  & Dec & (kpc)  \\
\hline
PAndAS-01 & 23 57 12.03 & +43 33 08.28 &       118.92 \\
PAndAS-02 & 23 57 55.69 & +41 46 49.25 &       114.74 \\
PAndAS-03 & 00 03 56.41 & +40 53 19.20 &       100.00 \\
PAndAS-04 & 00 04 42.93 & +47 21 42.47 &       124.62 \\
PAndAS-05 & 00 05 24.15 & +43 55 35.70 &       100.60 \\
PAndAS-06 & 00 06 11.95 & +41 41 20.97 &       93.66 \\
PAndAS-07 & 00 10 51.35 & +39 35 58.55 &       85.95 \\
PAndAS-08 & 00 12 52.45 & +38 17 47.86 &       88.26 \\
PAndAS-09 & 00 12 54.66 & +45 05 55.86 &       90.82 \\
PAndAS-10 & 00 13 38.66 & +45 11 11.13 &       90.00 \\
PAndAS-11 & 00 14 55.63 & +44 37 16.35 &       83.23 \\
PAndAS-12 & 00 17 40.08 & +43 18 39.02 &       69.21 \\
PAndAS-13 & 00 17 42.72 & +43 04 31.83 &       67.98 \\
PAndAS-14 & 00 20 33.88 & +36 39 34.46 &       86.20 \\
PAndAS-15 & 00 22 44.07 & +41 56 14.16 &       51.90 \\
PAndAS-16 & 00 24 59.92 & +39 42 13.11 &       50.81 \\
PAndAS-17 & 00 26 52.20 & +38 44 58.11 &       53.93 \\
PAndAS-18 & 00 28 23.26 & +39 55 04.86 &       41.55 \\
PAndAS-19 & 00 30 12.22 & +39 50 59.27 &       37.87 \\
PAndAS-20 & 00 31 23.74 & +41 59 20.12 &       30.59 \\
PAndAS-21 & 00 31 27.52 & +39 32 21.84 &       37.68 \\
PAndAS-22 & 00 32 08.36 & +40 37 31.62 &       28.73 \\
PAndAS-23 & 00 33 14.13 & +39 35 15.93 &       33.74 \\
PAndAS-24 & 00 33 50.57 & +38 38 28.04 &       42.81 \\
PAndAS-25 & 00 34 06.15 & +43 15 06.65 &       34.79 \\
PAndAS-26 & 00 34 45.08 & +38 26 38.05 &       43.92 \\
PAndAS-27 & 00 35 13.53 & +45 10 37.85 &       56.58 \\
PAndAS-28 & 00 35 56.43 & +40 48 44.98 &       18.60 \\
PAndAS-29 & 00 36 09.08 & +40 08 09.85 &       23.04 \\
PAndAS-30 & 00 38 29.01 & +37 58 39.21 &       46.35 \\
PAndAS-31 & 00 39 59.79 & +43 03 19.67 &       25.38 \\
PAndAS-32 & 00 40 41.20 & +40 00 54.95 &       17.94 \\
PAndAS-33 & 00 40 57.35 & +38 38 10.24 &       36.28 \\
PAndAS-34 & 00 41 18.04 & +42 46 16.51 &       20.85 \\
PAndAS-35 & 00 43 09.36 & +40 36 38.23 &       9.07 \\
PAndAS-36 & 00 44 45.57 & +43 26 34.79 &       30.14 \\
PAndAS-37 & 00 48 26.53 & +37 55 42.14 &       48.06 \\
PAndAS-38 & 00 49 45.67 & +47 54 33.12 &       92.33 \\
PAndAS-39 & 00 50 36.22 & +42 31 49.29 &       26.40 \\
PAndAS-40 & 00 50 43.80 & +40 03 30.20 &       26.51 \\
PAndAS-41 & 00 53 39.58 & +42 35 14.98 &       33.09 \\
PAndAS-42 & 00 56 38.04 & +39 40 25.93 &       42.18 \\
PAndAS-43 & 00 56 38.80 & +42 27 17.77 &       38.92 \\
PAndAS-44 & 00 57 55.89 & +41 42 57.01 &       39.35 \\
PAndAS-45 & 00 58 37.96 & +41 57 11.48 &       41.66 \\
PAndAS-46 & 00 58 56.40 & +42 27 38.29 &       44.31 \\
PAndAS-47 & 00 59 04.78 & +42 22 35.06 &       44.26 \\
PAndAS-48 & 00 59 28.26 & +31 29 10.64 &       141.34 \\
PAndAS-49 & 01 00 50.07 & +42 18 13.25 &       48.21 \\
PAndAS-50 & 01 01 50.66 & +48 18 19.22 &       106.68 \\
PAndAS-51 & 01 02 06.61 & +42 48 06.64 &       53.42 \\
PAndAS-52 & 01 12 47.01 & +42 25 24.87 &       78.05 \\
PAndAS-53 & 01 17 58.41 & +39 14 53.20 &       95.88 \\
PAndAS-54 & 01 18 00.14 & +39 16 59.93 &       95.79 \\
PAndAS-55 & 01 19 20.41 & +46 03 11.52 &       111.50 \\
PAndAS-56 & 01 23 03.53 & +41 55 11.02 &       103.34 \\
PAndAS-57 & 01 27 47.51 & +40 40 47.20 &       116.41 \\
PAndAS-58 & 01 29 02.16 & +40 47 08.66 &       119.42 \\
PAndAS-59 & 00 36 29.53 & +40 38 16.83 &       18.28 \\
PAndAS-Cand-01 & 00 44 58.35 & +40 21 37.92 &       13.70 \\
PAndAS-Cand-02 & 01 07 53.88 & +48 22 41.79 &       114.60 \\
\hline
\end{tabular}
\end{minipage}
\end{table}

\begin{figure*}
 \centering
 \includegraphics[width=140mm]{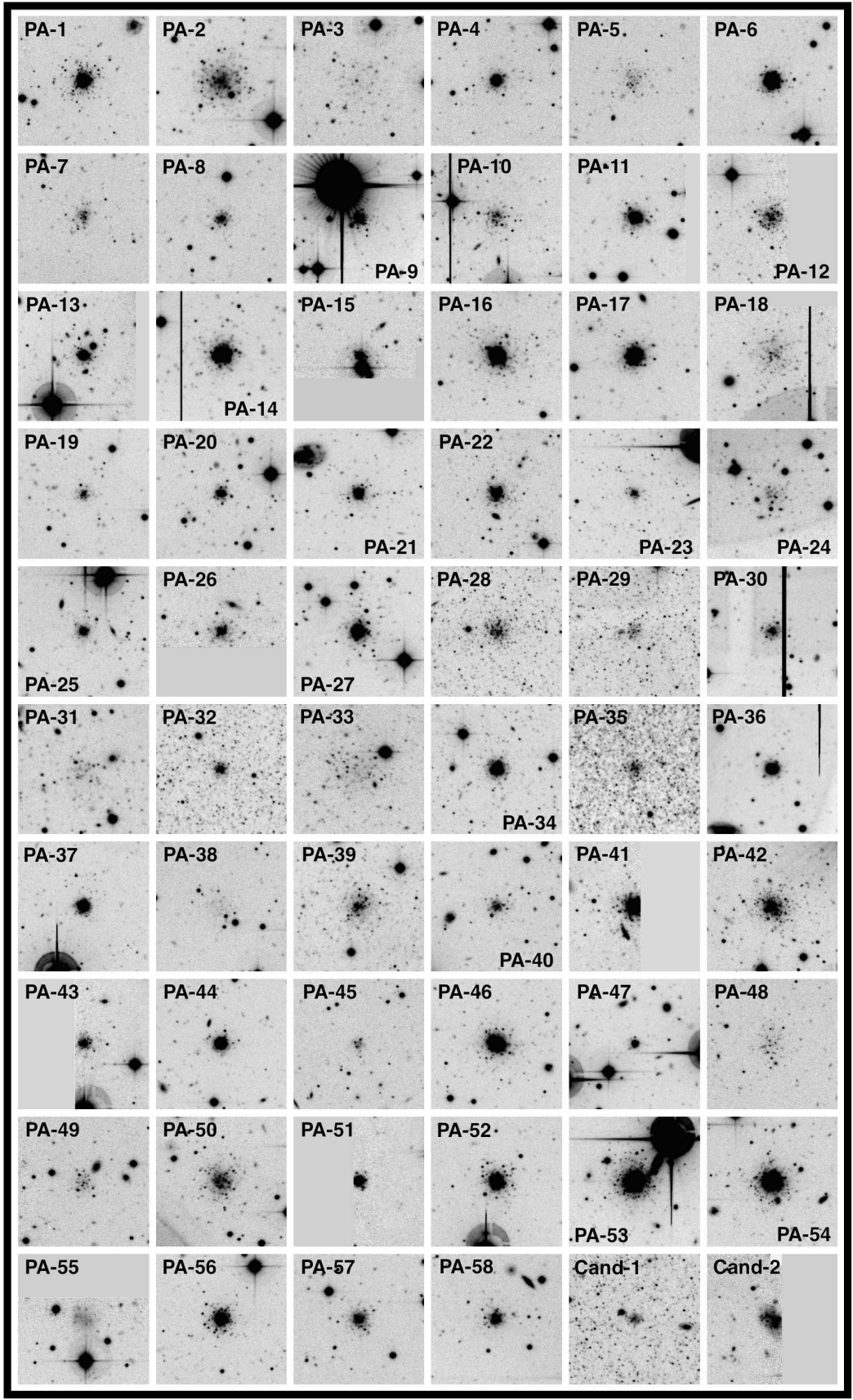}
 \vspace{2pt}
 \caption[g-band images of the new clusters from the MC data]
 {MegaCam $g$-band thumbnail images of our new M31 halo GCs.  Each image is $1\arcmin \times 1\arcmin$ in size, with north to the top and east to the left.  PA-59 is shown in Figure \ref{f:PA59}.}\label{Fi:mosaic}
\end{figure*}

The identity and location of each of these new objects are listed in Table \ref{tab:locations}, and g-band thumbnail images are shown in Figure \ref{Fi:mosaic}. The thumbnail images clearly reveal the unambiguous classification of each catalogued object as a GC, highlighting the quality of the data and the reason why our search turned up so few candidates with indeterminate classification. Of the two such candidates in our sample, Cand-01 is a faint object set against a relatively dense stellar background that hinders discrimination between its identity as a cluster or a distant galaxy, while Cand-02 is an extended object (cluster or distant galaxy) largely cut-off at the edge of an image. These are listed in Table \ref{tab:locations} but not included in our subsequent analysis. We aim to obtain follow-up observations of these objects to clarify their status. 

Note that with the 40 GCs presented in Hux08, we have found a total of $\sim$100 new GCs in the outer halo of M31.

For completeness we note that a number of the clusters listed in Table \ref{tab:locations} have formed the basis of previous studies -- specifically those by \citet{Mackeyetal10a} and \citet{Veljanoskietal13a} who investigated the ensemble properties of M31 halo GCs including subsamples from the present list, and the works by \citet{Mackeyetal13a,Mackeyetal13b} who studied specific objects (PA-48, and PA-7 and PA-8, respectively).

Our new catalogue represents the first detailed census of GCs across the full M31 halo, greatly extending the work of Hux08. The vast majority of our discoveries (53 of 59) lie in the outskirts of M31, at projected radii $R_{\rm proj} > 25$ kpc. Of these, a substantial fraction lie at distances that were completely unexplored prior to our CFHT campaign: our catalogue contains 21 clusters beyond $R_{\rm proj} = 80$ kpc, of which 11 sit outside $R_{\rm proj} = 100$ kpc. Indeed we effectively find GCs out to distances commensurate with the edge of the PAndAS footprint, confirming previous suggestions that the M31 cluster system is very extended \citep[e.g.,][]{Mackeyetal10b} and suggesting that additional GCs may be found at even larger radii \citep[see also][]{dtz:13}. Combined with previous discoveries, mostly from Hux08, we now know of 91 M31 GCs lying outside $R_{\rm proj} > 25$ kpc, which includes 12 at distances larger than $R_{\rm proj} = 100$ kpc.

These observations stand in stark contrast to the halo GC population in the Milky Way, in which there are only $\approx 13$ objects known at Galactocentric radii larger than $30-35$ kpc (corresponding to an average projected radius of $\sim 25$ kpc for random viewing angles), and in which the most distant known member sits at a Galactocentric distance of $\approx 120$ kpc (corresponding to an average projected distance of $\sim 95$ kpc for random viewing angles). While the disparity in the number of GCs in the Milky Way and M31 within $R_{\rm proj} \approx 25$ kpc is roughly 3:1 in favour of M31, our new catalogue reveals that outside this radius it is more like 7:1 in favour of M31. We explore the differences between these two GC systems in more detail in Section \ref{analysis}.

The photometric properties (luminosities, colours and sizes) of our new GC sample are derived below in Section \ref{photometry}. However, the excellent quality of PAndAS imaging also allowed us to examine and resolve the identity of many candidate clusters previously identified in the literature, and we first turn our attention to these.

\section{Updates to Published Catalogues}
\label{other_catalogue_updates}

There are two primary sources of candidate M31 halo clusters -- the Revised Bologna Catalogue \citep[RBC;][]{galleti:04}, and a recent search for M31 GCs in the Sloan Digital Sky Survey (SDSS) by \citet{dtz:13}.

\subsection{Revised Bologna Catalogue}
\label{RBCupdate}
The RBC is the main repository for information on the M31 GC system. It contains lists of confirmed GCs (classes $1$ and $8$ in the catalogue) and candidate GCs (classes $2$ and $3$), as well as a few H{\sc ii} regions (class $5$), and compilations of objects once suspected to be GCs but subsequently revealed as background galaxies (class $4$) or foreground stars (class $6$). The identities of these contaminants are retained in the RBC in order to avoid mis-classification in future GC surveys.

We inspected $1.5\arcmin \times 1.5\arcmin$ PAndAS thumbnails of all objects listed in Version 5 of the RBC (released in August 2012) as having projected galactocentric radii larger than $R_{\rm proj} \sim 15$ kpc. Inside this radius the strong and variable background due to the M31 disk means that even with our high quality PAndAS imaging it is frequently impossible to establish a reliable target classification\footnote{Note that this radius is smaller than the inner radius of our uniform survey ($R_{\rm proj} = 25$ kpc) as here we are not trying to discover new GCs, but rather establish classifications for objects for which we already have positions.}. Overall there are $523$ objects with $R_{\rm proj} \geq 15$ kpc in the RBC V5, of which $497$ have PAndAS imaging. The missing $26$ entries typically fall into small gaps in the coverage resulting from the inter-row CCD spaces on the MegaCam array (the spaces between individual CCDs on a given row were covered by the PAndAS dither pattern) or imperfect tiling of the PAndAS mosaic, although a couple sit outside the survey footprint with $R_{\rm proj} \ga 150$ kpc.

To avoid, as far as possible, prior knowledge introducing bias into our classifications, we employed a blind inspection methodology. One of us (ADM) generated thumbnails for all targets, randomised the order, and supplied the images only, with no supplementary information, to APH for classification. Once the inspection process was complete, the results were returned to ADM for analysis.

The original RBC classifications for the $497$ objects we inspected broke down as follows: $72$ GCs, $141$ GC candidates, $166$ background galaxies, $116$ foreground stars, and $2$ H{\sc ii} regions. We confirmed that all $282$ of the contaminant objects (galaxies and stars) were correctly identified as such. Of more interest are the GCs and GC candidates, and we were able to greatly improve classifications for these targets. 

\begin{table*}
\centering
\caption{Updated globular cluster classifications in the Revised Bologna Catalogue V5.}
\begin{minipage}{112mm}
\begin{tabular}{@{}lccccc}
\hline \hline
Name in & \multicolumn{2}{c}{Position (J2000.0)} & $R_{\rm proj}$ & Previous & New \\
RBC V5 & RA & Dec & (kpc) & Classification$^a$ & Classification$^a$ \\
\hline
\multicolumn{6}{l}{\hspace{-2.2mm}{\bf Promoted GCs + H{\sc ii}}} \\
PAndAS-59 & $00\,36\,29.35$ & $+40\,38\,16.7$ & $18.29$ & $-$ & $1$ \\
SH06 & $00\,39\,19.05$ & $+40\,21\,58.0$ & $15.19$ & $2$ & $5\,(1?)$ \\
B270D & $00\,45\,49.22$ & $+41\,01\,49.3$ & $8.57$ & $2$ & $1$ \\
SK213C & $00\,46\,58.77$ & $+42\,17\,45.3$ & $17.71$ & $2$ & $1$ \\
SK255B & $00\,49\,03.02$ & $+41\,54\,57.8$ & $18.39$ & $2$ & $1$\vspace{1mm} \\
\multicolumn{6}{l}{\hspace{-2.2mm}{\bf Demoted GCs}} \\
SK002A & $00\,36\,34.99$ & $+41\,01\,08.0$ & $16.20$ & $1$ & $6$ \\
SK004A & $00\,38\,01.35$ & $+42\,04\,06.4$ & $16.25$ & $1$ & $6$ \\
BA11 & $00\,48\,45.59$ & $+42\,23\,37.7$ & $21.71$ & $1$ & $4+6$ \\
\hline
\label{t:rbc1}
\end{tabular} 
\medskip
\vspace{-7mm}
\\
$^a$ Classes: $1,8$ = GC; $2,3$ = candidate GC; $4$ = galaxy; $5$ = H{\sc ii} region; $6$ = star.\\
$^b$ Classified as a candidate in previous versions of the RBC -- see text.
\end{minipage}
\end{table*}
  
\begin{figure*}
\begin{center}
\includegraphics[width=136mm]{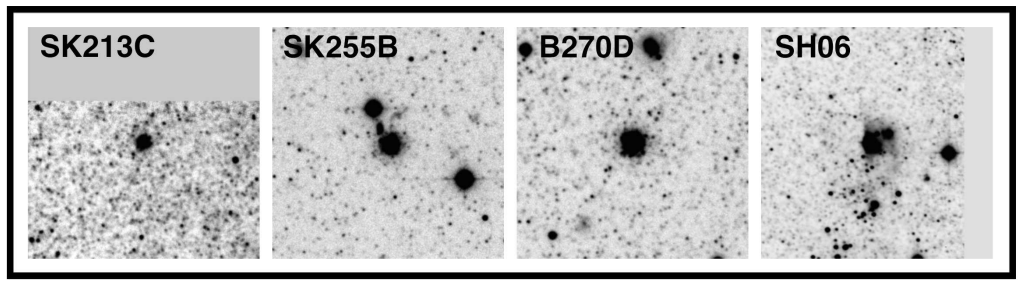}
\caption{PAndAS $g$-band thumbnails for the three new globular clusters uncovered in the RBC V5, plus SH06 (see text). Each thumbnail is $1\arcmin \times 1\arcmin$ in size, with north to the top and east to the left.}
\label{f:newGCs}
\end{center}
\end{figure*}

\begin{figure}
\begin{center}
\includegraphics[width=75mm]{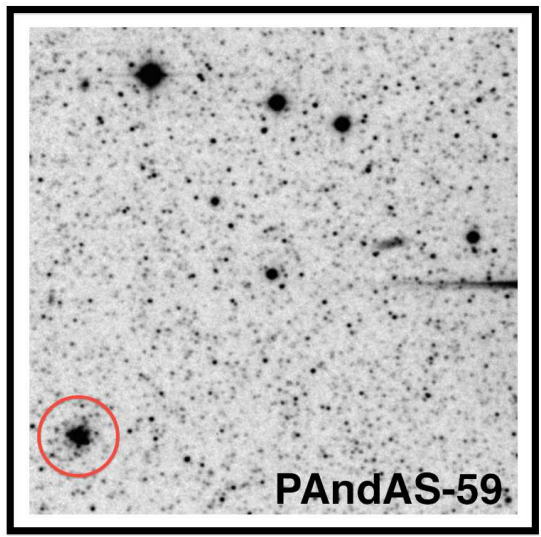}
\caption{PAndAS $g$-band thumbnail for our serendipitous discovery PAndAS-59 (circled). The object at the centre of the field is SK014B, correctly classified in the RBC as a star. The thumbnail is $1.5\arcmin \times 1.5\arcmin$, with north to the top and east to the left.}
\label{f:PA59}
\end{center}
\end{figure}
  
We found that two of the candidates were in fact genuine GCs. These objects are listed in Table \ref{t:rbc1} and their $g$-band thumbnails displayed in Figure \ref{f:newGCs}: SK213C and SK255B. Both are located within the 25 kpc inner limit of our main survey, which was why they were not identified as part of that search. We also uncovered one particularly interesting GC candidate -- SH06, which consists of a compact luminous source surrounded by nebulosity that is quite evident in the $g$-band imaging (see Figure \ref{f:newGCs}) but virtually invisible in the $i$-band. This is suggestive of a massive young star cluster still embedded in gas, sitting at $ \sim 15$ kpc from the M31 centre. However, because we cannot be absolutely certain that there is a cluster present, we conservatively classify this object as a H{\sc ii} region in Table \ref{t:rbc1}. 

While inspecting the object SK014B, which is correctly classified in the RBC V5 as a star, we noticed a small cluster nearby which does not appear anywhere in the RBC.  We therefore believe this to be a new discovery, which we name PAndAS-59. We include PA-59 in Table \ref{t:rbc1} and display the discovery thumbnail in Figure \ref{f:PA59}. That we identified this object serendipitously in our sample of small thumbnail images suggests that a full search of the inner M31 halo, between $\sim 15-25$ kpc, may be quite fruitful -- although we reiterate the caveat that the increased crowding would adversely affect the completeness of any such survey.

To our sample of new RBC GCs we also add B270D. At $R_{\rm proj} = 8.57$ kpc, this object sits well inside both our inner PAndAS  search radius, and our inner RBC inspection radius. In V5 of the RBC it is classified as a candidate object; we uncovered its cluster status by chance. It is listed in Table \ref{t:rbc1} and displayed in Figure \ref{f:newGCs}.

\begin{figure*}
\begin{center}
\includegraphics[width=102mm]{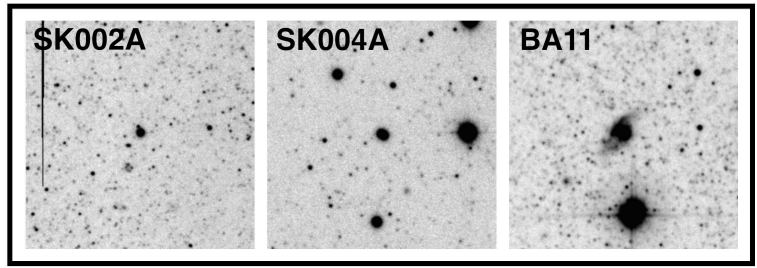}
\caption{PAndAS $g$-band thumbnails for objects mis-classified as globular clusters in the RBC V5. Each thumbnail is $1\arcmin \times 1\arcmin$ in size, with north to the top and east to the left. SK002A is a star; SK004A is two barely-separated stars; and BA11 is a star superimposed on a background galaxy.}
\label{f:notGCs}
\end{center}
\end{figure*}

In addition to confirming several new globular clusters among RBC candidates, we also found three ``confirmed'' RBC GCs which were misclassified stars or galaxies. These objects -- SK002A, SK004A, and BA11 -- are listed in Table \ref{t:rbc1} and their thumbnails shown in Figure \ref{f:notGCs}.

\begin{table*}
\centering
\caption{Updated classifications for globular cluster candidates listed in the Revised Bologna Catalogue V5.}
\begin{tabular}{@{}lccccclcccc}
\hline \hline
Name in & \multicolumn{2}{c}{Position (J2000.0)} & $R_{\rm proj}$ & New & & Name in & \multicolumn{2}{c}{Position (J2000.0)} & $R_{\rm proj}$ & New \\
RBC V5 & RA & Dec & (kpc) & Class$^a$ & & RBC V5 & RA & Dec & (kpc) & Class$^a$ \\
\hline
SK001C & $00\,33\,13.080$ & $+40\,05\,26.00$ & $29.47$ & $4$ & \hspace{6mm} & SK019B & $00\,37\,33.470$ & $+40\,05\,28.70$ & $20.96$ & $4$ \\
SK002C & $00\,33\,15.820$ & $+40\,00\,24.70$ & $30.03$ & $4$ & & SK020B & $00\,37\,35.680$ & $+40\,35\,14.40$ & $16.22$ & $4$ \\
SK001B & $00\,33\,23.070$ & $+40\,04\,40.70$ & $29.21$ & $4$ & & SK048C & $00\,37\,37.200$ & $+40\,05\,39.60$ & $20.83$ & $4$ \\
SK002B & $00\,33\,32.200$ & $+39\,51\,32.80$ & $30.70$ & $4$ & & SK049C & $00\,37\,37.270$ & $+41\,54\,04.40$ & $15.67$ & $4$ \\
SK004C & $00\,33\,34.960$ & $+40\,08\,16.30$ & $28.32$ & $4$ & & SK050C & $00\,37\,41.270$ & $+40\,04\,42.90$ & $20.89$ & $4$ \\
SK003B & $00\,33\,37.030$ & $+39\,40\,59.00$ & $32.14$ & $4$ & & SK051C & $00\,37\,41.790$ & $+40\,05\,18.00$ & $20.77$ & $4$ \\
SK005C & $00\,33\,38.040$ & $+39\,35\,35.70$ & $32.96$ & $4$ & & SK022B & $00\,37\,54.310$ & $+40\,17\,26.70$ & $18.31$ & $4$ \\
SK006C & $00\,33\,46.110$ & $+39\,48\,36.60$ & $30.67$ & $4$ & & SK053C & $00\,38\,00.760$ & $+42\,02\,56.90$ & $16.10$ & $4$ \\
SK007C & $00\,33\,54.630$ & $+39\,34\,36.60$ & $32.61$ & $4$ & & SK054C & $00\,38\,06.100$ & $+40\,24\,30.20$ & $16.80$ & $6$ \\
B133D & $00\,34\,10.994$ & $+39\,50\,50.27$ & $29.52$ & $4$ & & SK058C & $00\,38\,48.400$ & $+40\,03\,01.20$ & $19.53$ & $4$ \\
BH01$^b$ & $00\,34\,11.480$ & $+39\,23\,59.10$ & $33.90$ & $-$ & & SK066C & $00\,39\,15.190$ & $+42\,22\,50.70$ & $17.59$ & $6$ \\
SK009C & $00\,34\,12.200$ & $+40\,06\,29.70$ & $27.22$ & $4$ & & B186D & $00\,40\,02.258$ & $+39\,23\,12.11$ & $26.68$ & $4$ \\
SK010C & $00\,34\,26.850$ & $+39\,54\,05.60$ & $28.51$ & $4$ & & SK073C & $00\,40\,04.300$ & $+40\,14\,10.70$ & $15.72$ & $4$ \\
B411 & $00\,34\,30.808$ & $+41\,33\,44.09$ & $21.46$ & $4$ & & B188D & $00\,40\,14.038$ & $+39\,41\,30.82$ & $22.52$ & $4$ \\
SK004B & $00\,34\,34.200$ & $+40\,02\,49.40$ & $26.98$ & $4$ & & B191D & $00\,40\,17.893$ & $+42\,25\,23.98$ & $16.95$ & $4$ \\
SK011C & $00\,34\,51.160$ & $+39\,55\,33.10$ & $27.50$ & $4$ & & SK090C & $00\,40\,53.060$ & $+40\,00\,43.30$ & $17.84$ & $4$ \\
B412 & $00\,34\,55.281$ & $+41\,32\,26.49$ & $20.38$ & $4$ & & B460 & $00\,41\,54.817$ & $+39\,35\,25.51$ & $23.05$ & $4$ \\
SK012C & $00\,35\,08.810$ & $+40\,07\,32.60$ & $25.13$ & $4$ & & SK104C & $00\,42\,03.040$ & $+40\,03\,48.80$ & $16.58$ & $4$ \\
SK013C & $00\,35\,09.240$ & $+40\,05\,39.80$ & $25.38$ & $4$ & & SK110C & $00\,42\,33.090$ & $+40\,04\,53.60$ & $16.24$ & $4$ \\
B413 & $00\,35\,13.001$ & $+41\,29\,07.81$ & $19.51$ & $4$ & & B225D & $00\,43\,13.440$ & $+40\,01\,14.58$ & $17.11$ & $6$ \\
BA22 & $00\,35\,13.608$ & $+39\,45\,37.16$ & $28.40$ & $4$ & & B233D & $00\,43\,41.311$ & $+39\,36\,45.96$ & $22.78$ & $4$ \\
SK014C & $00\,35\,14.860$ & $+39\,41\,40.00$ & $29.03$ & $4$ & & SK136C & $00\,44\,04.430$ & $+40\,05\,19.60$ & $16.50$ & $6$ \\
SK015C & $00\,35\,20.470$ & $+39\,35\,04.10$ & $30.02$ & $4$ & & SK160C & $00\,44\,54.430$ & $+40\,06\,44.10$ & $16.78$ & $4$ \\
SK016C & $00\,35\,22.000$ & $+41\,49\,47.40$ & $20.35$ & $4$ & & SK205B & $00\,45\,33.250$ & $+40\,17\,08.40$ & $15.29$ & $4$ \\
SK017C & $00\,35\,28.440$ & $+39\,32\,25.10$ & $30.27$ & $6$ & & SK193C & $00\,45\,49.970$ & $+40\,05\,09.10$ & $18.05$ & $4$ \\
SK018C & $00\,35\,29.320$ & $+41\,42\,33.30$ & $19.51$ & $4$ & & SK196C & $00\,45\,51.580$ & $+40\,04\,43.80$ & $18.17$ & $4$ \\
B134D & $00\,35\,30.298$ & $+40\,44\,24.84$ & $20.01$ & $4$ & & SK214B & $00\,45\,54.060$ & $+39\,56\,46.80$ & $19.86$ & $6$ \\
SK006B & $00\,35\,34.240$ & $+41\,11\,53.00$ & $18.45$ & $4$ & & SK197C & $00\,45\,57.630$ & $+40\,17\,09.50$ & $15.82$ & $4$ \\
SK007B & $00\,35\,45.260$ & $+39\,39\,21.30$ & $28.57$ & $4$ & & SK200C & $00\,46\,06.090$ & $+40\,22\,26.00$ & $15.01$ & $4$ \\
SK020C & $00\,35\,49.740$ & $+41\,50\,02.40$ & $19.28$ & $4$ & & SK221B & $00\,46\,19.240$ & $+40\,23\,42.00$ & $15.12$ & $6$ \\
SK021C & $00\,35\,50.830$ & $+39\,36\,00.80$ & $29.02$ & $4$ & & B281D & $00\,46\,22.279$ & $+40\,18\,08.00$ & $16.22$ & $6$ \\
SK022C & $00\,35\,51.760$ & $+40\,54\,11.60$ & $18.40$ & $4$ & & SK204C & $00\,46\,22.920$ & $+40\,20\,42.20$ & $15.76$ & $4$ \\
SK023C & $00\,35\,53.100$ & $+41\,51\,23.70$ & $19.27$ & $4$ & & SK223B & $00\,46\,32.880$ & $+40\,06\,36.50$ & $18.67$ & $6$ \\
SK024C & $00\,35\,53.830$ & $+41\,43\,42.60$ & $18.60$ & $4$ & & B488$^c$ & $00\,46\,34.287$ & $+42\,11\,42.78$ & $15.99$ & $5$ \\
SK025C & $00\,35\,54.220$ & $+41\,46\,53.80$ & $18.84$ & $4$ & & B489 & $00\,46\,36.386$ & $+40\,00\,26.86$ & $19.95$ & $4$ \\
SK008B & $00\,35\,58.150$ & $+39\,37\,35.50$ & $28.54$ & $4$ & & SH21 & $00\,46\,37.308$ & $+39\,23\,57.85$ & $27.49$ & $6$ \\
SK009B & $00\,36\,00.230$ & $+40\,56\,19.20$ & $17.92$ & $4$ & & B291D & $00\,46\,41.270$ & $+40\,03\,02.00$ & $19.55$ & $6$ \\
SK010B & $00\,36\,01.700$ & $+39\,48\,50.20$ & $26.45$ & $4$ & & B293D & $00\,46\,48.097$ & $+40\,02\,21.72$ & $19.84$ & $6$ \\
SK011B & $00\,36\,02.020$ & $+41\,14\,43.40$ & $17.23$ & $4$ & & B390 & $00\,46\,51.632$ & $+40\,23\,46.90$ & $16.00$ & $6$ \\
SK026C & $00\,36\,05.610$ & $+39\,58\,04.90$ & $24.77$ & $4$ & & SK231B & $00\,47\,14.110$ & $+40\,22\,23.20$ & $16.89$ & $6$ \\
SK029C & $00\,36\,22.260$ & $+39\,52\,04.50$ & $25.30$ & $4$ & & BA28 & $00\,47\,14.220$ & $+42\,21\,42.20$ & $18.82$ & $4$ \\
B139D & $00\,36\,24.679$ & $+39\,45\,07.43$ & $26.47$ & $6$ & & SK232B & $00\,47\,14.430$ & $+40\,25\,38.80$ & $16.36$ & $4$ \\
SK030C & $00\,36\,27.360$ & $+41\,35\,14.00$ & $16.67$ & $4$ & & DAO93 & $00\,47\,46.178$ & $+42\,44\,55.88$ & $23.92$ & $4$ \\
SK031C & $00\,36\,31.430$ & $+42\,06\,24.60$ & $19.56$ & $4$ & & DAO94 & $00\,47\,54.399$ & $+42\,44\,01.58$ & $23.94$ & $4$ \\
SK013B & $00\,36\,31.700$ & $+41\,11\,41.30$ & $15.99$ & $4$ & & BA10 & $00\,47\,56.286$ & $+42\,28\,43.73$ & $21.18$ & $4$ \\
SK032C & $00\,36\,33.360$ & $+41\,30\,03.10$ & $16.16$ & $4$ & & SK222C & $00\,47\,59.480$ & $+41\,54\,13.00$ & $15.98$ & $6$ \\
B142D & $00\,36\,33.831$ & $+41\,09\,07.96$ & $15.95$ & $4$ & & SK238B & $00\,48\,01.660$ & $+41\,49\,56.90$ & $15.56$ & $6$ \\
B144D & $00\,36\,36.647$ & $+41\,37\,03.65$ & $16.40$ & $6$ & & SK223C & $00\,48\,04.610$ & $+40\,08\,27.00$ & $20.72$ & $4$ \\
SK033C & $00\,36\,37.890$ & $+42\,14\,46.20$ & $20.51$ & $4$ & & SH24 & $00\,48\,15.545$ & $+42\,25\,17.12$ & $21.11$ & $6$ \\
SK034C & $00\,36\,43.420$ & $+39\,34\,56.20$ & $27.87$ & $4$ & & SK240B & $00\,48\,24.140$ & $+40\,06\,43.10$ & $21.58$ & $4$ \\
SK035C & $00\,36\,46.660$ & $+41\,26\,23.90$ & $15.47$ & $4$ & & SK243B & $00\,48\,27.190$ & $+42\,02\,43.50$ & $18.04$ & $4$ \\
SK036C & $00\,36\,47.200$ & $+40\,04\,09.50$ & $22.52$ & $4$ & & SK225C & $00\,48\,31.340$ & $+42\,01\,05.50$ & $17.97$ & $4$ \\
SK037C & $00\,36\,49.190$ & $+39\,39\,43.70$ & $26.82$ & $4$ & & SK249B & $00\,48\,32.960$ & $+42\,02\,45.00$ & $18.24$ & $4$ \\
SK039C & $00\,37\,00.980$ & $+39\,33\,27.60$ & $27.74$ & $6$ & & SK252B & $00\,48\,41.130$ & $+41\,31\,54.60$ & $15.66$ & $6$ \\
SK040C & $00\,37\,03.400$ & $+41\,33\,22.10$ & $15.08$ & $6$ & & SK227C & $00\,48\,44.050$ & $+42\,15\,48.80$ & $20.45$ & $4$ \\
SK041C & $00\,37\,05.080$ & $+40\,01\,06.30$ & $22.52$ & $4$ & & B504 & $00\,48\,45.168$ & $+40\,08\,45.94$ & $21.87$ & $4$ \\
SK042C & $00\,37\,06.280$ & $+41\,44\,48.50$ & $15.82$ & $4$ & & SK228C & $00\,48\,46.490$ & $+41\,46\,45.80$ & $16.94$ & $4$ \\
SK043C & $00\,37\,09.100$ & $+39\,49\,10.40$ & $24.56$ & $6$ & & DAO99 & $00\,48\,48.314$ & $+42\,32\,45.20$ & $23.29$ & $4$ \\
SK046C & $00\,37\,28.930$ & $+41\,55\,01.90$ & $16.09$ & $4$ & & B334D & $00\,48\,54.848$ & $+39\,35\,56.07$ & $27.92$ & $4$ \\
SK017B & $00\,37\,30.440$ & $+40\,36\,43.80$ & $16.22$ & $4$ & & SK256B & $00\,49\,05.380$ & $+41\,57\,38.30$ & $18.78$ & $4$ \\
SK018B & $00\,37\,30.820$ & $+40\,18\,23.90$ & $18.86$ & $4$ & & SK229C & $00\,49\,11.040$ & $+41\,57\,53.10$ & $19.01$ & $4$ \\
\hline
\label{t:rbc2}
\end{tabular}
\end{table*}

\addtocounter{table}{-1}
\begin{table*}
\centering
\caption{Continued.}
\begin{minipage}{167mm}
\begin{tabular}{@{}lccccclcccc}
\hline \hline
Name in & \multicolumn{2}{c}{Position (J2000.0)} & $R_{\rm proj}$ & New & & Name in & \multicolumn{2}{c}{Position (J2000.0)} & $R_{\rm proj}$ & New \\
RBC V5 & RA & Dec & (kpc) & Class$^a$ & & RBC V5 & RA & Dec & (kpc) & Class$^a$ \\
\hline
SK257B & $00\,49\,15.210$ & $+41\,01\,29.40$ & $17.09$ & $6$ & \hspace{6mm} & B346D & $00\,50\,03.750$ & $+40\,37\,39.28$ & $20.84$ & $4$ \\
B338D & $00\,49\,15.765$ & $+40\,46\,23.47$ & $18.14$ & $4$ & & SK258B & $00\,50\,17.460$ & $+42\,06\,42.60$ & $22.45$ & $4$ \\
B339D & $00\,49\,17.493$ & $+40\,45\,06.58$ & $18.32$ & $4$ & & B348D & $00\,50\,19.219$ & $+40\,58\,02.78$ & $19.95$ & $4$ \\
DAO104 & $00\,49\,21.347$ & $+42\,16\,16.60$ & $21.73$ & $4$ & & B511 & $00\,50\,43.418$ & $+40\,11\,13.39$ & $25.43$ & $4$ \\
SK232C & $00\,49\,25.630$ & $+42\,06\,06.70$ & $20.52$ & $4$ & & B512 & $00\,50\,46.324$ & $+39\,53\,19.91$ & $28.13$ & $4$ \\
B340D & $00\,49\,29.174$ & $+41\,04\,32.10$ & $17.56$ & $4$ & & B513 & $00\,50\,47.806$ & $+41\,25\,46.27$ & $20.79$ & $4$ \\
B506 & $00\,49\,34.905$ & $+40\,00\,28.94$ & $24.74$ & $4$ & & G355 & $00\,51\,33.740$ & $+39\,57\,35.81$ & $29.06$ & $4$ \\
SK233C & $00\,49\,35.650$ & $+42\,11\,42.80$ & $21.58$ & $4$ & & SH25 & $00\,52\,04.054$ & $+41\,35\,05.85$ & $24.29$ & $4$ \\
B345D$^d$ & $00\,49\,52.554$ & $+40\,53\,10.10$ & $19.12$ & $6$ & &  \\
\hline
\end{tabular}
\medskip
\vspace{-2mm}
\\
$^a$ All objects were originally classified as cluster candidates (class $2$ or $3$) except for B488 (class $5$).\\
$^b$ No object is visible at the coordinates specified for BH01. This is a very faint candidate from archival {\it HST} imaging.\\
$^c$ We retain the classification of B488 as a H{\sc ii} region but note there may also be a young cluster at this location.\\
$^d$ B345D appears to be a star superposed on a galaxy, so could also be classed as $4$.
\end{minipage}
\end{table*}
 
\begin{figure*}
\begin{center}
\includegraphics[width=170mm]{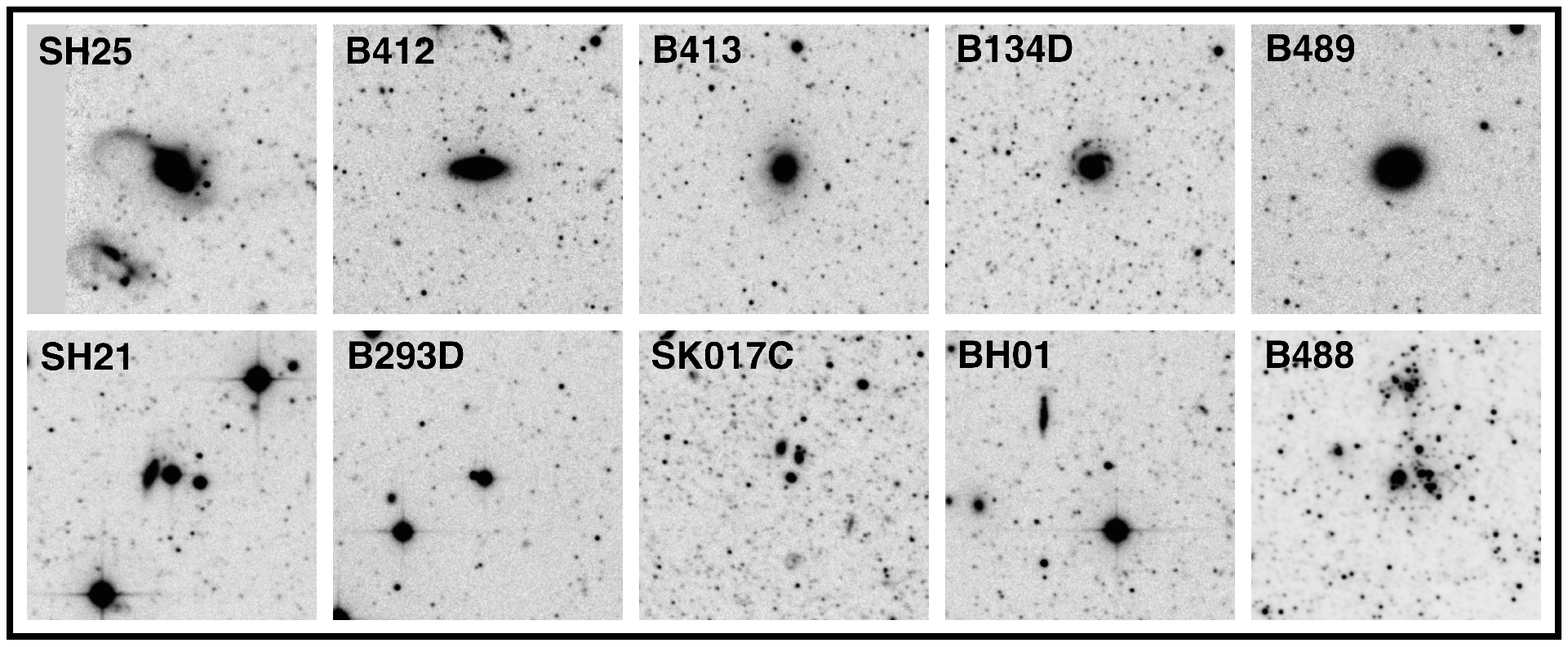}
\caption{PAndAS $g$-band thumbnails for representative examples of objects classified as globular cluster candidates in the RBC V5, that are galaxies (upper row) or stars (lower row, left three panels). In addition we include on the lower row images for BH01, for which no object is visible at the listed coordinates, and B488, which is a H{\sc ii} region that may also contain a dispersed young cluster. Each thumbnail is $1\arcmin \times 1\arcmin$ in size, with north to the top and east to the left.}
\label{f:stargal}
\end{center}
\end{figure*}

All but one of the remaining $138$ objects originally listed as GC candidates in the RBC V5 are either foreground stars 
($25$ objects) or background galaxies ($112$ objects). These are listed in Table \ref{t:rbc2}, and a few representative examples are
displayed in Figure \ref{f:stargal}. For the last candidate, BH01, we could not find any object at the listed coordinates. This is likely
because BH01 is a very faint object identified from {\it HST} WFPC2 imaging \citep{barmby:01}. The
thumbnail for this target is also displayed in Figure \ref{f:stargal}.  

Finally, of the two H{\sc ii} regions listed in our sample, we found no compelling reason to alter the classification of one (DAO88), while at the coordinates of the second, B488, we found a dispersed sample of luminous blue stars and a small amount of nebulosity. This is consistent with its classification as a H{\sc ii} region; however we note
that there may possibly also be a young cluster at this location. A thumbnail for this object is shown in Figure \ref{f:stargal}.

In summary, we inspected PAndAS thumbnails for $497$ objects listed at $R_{\rm proj} \geq 15$ kpc in the RBC V5. Of these, $141$ were 
originally classified as GC candidates; we were able to reclassify these as genuine GCs ($2$ objects) plus a H{\sc ii} region with a possible embedded
young massive cluster, foreground stars ($25$ objects), and background galaxies ($112$ objects), while in one case no object was visible. Of the $72$
targets originally listed as definite GCs, we confirmed $69$ but found that three were either foreground stars or background
galaxies. We did not change the classification of the $2$ H{\sc ii} regions in the RBC list, and we confirmed the identity of the $282$ objects
originally listed as contaminants. Finally, we added two more new GCs (B270D and PAndAS-59) located by chance as discussed above. 

\subsection{Candidates from SDSS}

\begin{table}
\centering
\caption{Confirmed globular clusters in the dTZZ13 catalogue.}
\begin{tabular}{@{}lcccc}
\hline \hline
Name in & PAndAS & \multicolumn{2}{c}{Position (J2000.0)} & $R_{\rm proj}$ \\
dTZZ13 & Name & RA & Dec & (kpc) \\
\hline
SDSS1 & $-$ & $00\,36\,01.8$ & $+40\,29\,50$ & $20.29$ \\
SDSS3 & $-$ & $00\,39\,13.1$ & $+41\,42\,08$ & $10.78$ \\
SDSS4 & PA-34 & $00\,41\,18.0$ & $+42\,46\,16$ & $20.84$ \\
SDSS6 & $-$ & $00\,42\,27.6$ & $+39\,55\,28$ & $18.39$ \\
SDSS8 & PA-39 & $00\,50\,36.3$ & $+42\,31\,50$ & $26.40$ \\
SDSS9 & PA-41 & $00\,53\,39.6$ & $+42\,35\,15$ & $33.09$ \\
SDSS11 & PA-46 & $00\,58\,56.4$ & $+42\,27\,38$ & $44.31$ \\
SDSS12 & PA-52 & $01\,12\,47.0$ & $+42\,25\,25$ & $78.05$ \\
SDSS15 & PA-56 & $01\,23\,03.5$ & $+41\,55\,11$ & $103.34$ \\
SDSS16 & PA-58 & $01\,29\,02.2$ & $+40\,47\,09$ & $119.42$ \\
C62 & PA-57 & $01\,27\,47.6$ & $+40\,40\,48$ & $116.41$ \\
\hline
\label{t:zinn1}
\end{tabular}
\end{table}

\begin{figure*}
\begin{center}
\includegraphics[width=102mm]{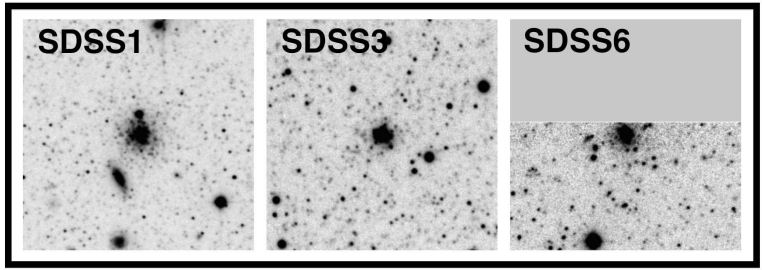}
\caption{PAndAS $g$-band thumbnails for the three confirmed non-PAndAS globular clusters in the SDSS catalogue of \citet{dtz:13}. Each thumbnail is $1\arcmin \times 1\arcmin$ in size, with north to the top and east to the left.}
\label{f:sdssGCs}
\end{center}
\end{figure*}

\begin{table}
\centering
\caption{Non-clusters in the dTZZ13 catalogue.}
\begin{tabular}{@{}lcccc}
\hline \hline
Name in & \multicolumn{2}{c}{Position (J2000.0)} & $R_{\rm proj}$ & Class \\
dTZZ13 & RA & Dec & (kpc) & \\
\hline
SDSS2 & $00\,38\,26.9$ & $+40\,12\,35$ & $18.25$ & $4$ \\
SDSS5 & $00\,41\,47.2$ & $+41\,44\,10$ & $6.83$ & $4$ \\
SDSS7 & $00\,47\,41.1$ & $+42\,04\,17$ & $16.72$ & $4$ \\
SDSS10 & $00\,55\,28.1$ & $+43\,59\,31$ & $49.06$ & $4$ \\
SDSS13 & $01\,16\,41.7$ & $+33\,19\,25$ & $142.35$ & $4$ \\
SDSS14 & $01\,22\,20.7$ & $+35\,11\,35$ & $134.73$ & $4$ \\
SDSS18 & $23\,49\,09.7$ & $+40\,27\,30$ & $138.73$ & $4$ \\
C2 & $00\,08\,19.0$ & $+34\,28\,07$ & $131.23$ & $4$ \\
C3 & $00\,08\,34.5$ & $+34\,37\,38$ & $129.14$ & $4$ \\
C14 & $00\,39\,32.3$ & $+40\,51\,17$ & $10.00$ & $4$ \\
C15 & $00\,40\,09.5$ & $+39\,55\,30$ & $19.55$ & $4$ \\
C16 & $00\,40\,14.0$ & $+39\,02\,33$ & $31.13$ & $4$ \\
C17 & $00\,40\,31.9$ & $+38\,11\,12$ & $42.52$ & $4$ \\
C18 & $00\,41\,38.9$ & $+37\,19\,34$ & $53.96$ & $4$ \\
C20 & $00\,42\,09.2$ & $+38\,56\,15$ & $31.90$ & $4$ \\
C22 & $00\,43\,03.7$ & $+32\,08\,37$ & $124.71$ & $4$ \\
C23 & $00\,43\,32.0$ & $+33\,10\,04$ & $110.73$ & $4$ \\
C24 & $00\,43\,44.3$ & $+31\,41\,24$ & $130.93$ & $4$ \\
C26 & $00\,44\,01.0$ & $+30\,42\,01$ & $144.48$ & $4$ \\
C27 & $00\,45\,40.2$ & $+37\,47\,11$ & $48.22$ & $4$ \\
C30 & $00\,48\,25.4$ & $+29\,16\,03$ & $164.77$ & $4$ \\
C31 & $00\,49\,33.1$ & $+34\,52\,00$ & $89.39$ & $4$ \\
C32 & $00\,49\,37.5$ & $+33\,44\,54$ & $104.45$ & $4$ \\
C33 & $00\,50\,22.5$ & $+41\,51\,35$ & $21.12$ & $4$ \\
C34 & $00\,51\,12.0$ & $+43\,33\,35$ & $37.88$ & $4$ \\
C36 & $00\,51\,32.6$ & $+41\,57\,24$ & $24.37$ & $4$ \\
C37 & $00\,51\,47.3$ & $+41\,37\,32$ & $23.68$ & $4$ \\
C39 & $00\,52\,34.6$ & $+43\,18\,25$ & $37.33$ & $4$ \\
C40 & $00\,54\,06.3$ & $+29\,55\,18$ & $158.23$ & $4$ \\
C41 & $01\,00\,12.7$ & $+34\,00\,43$ & $109.83$ & $4$ \\
C45 & $01\,05\,43.0$ & $+30\,56\,43$ & $154.59$ & $4$ \\
C47 & $01\,06\,14.8$ & $+34\,01\,15$ & $117.64$ & $4$ \\
C48 & $01\,06\,40.5$ & $+32\,29\,59$ & $136.44$ & $4$ \\
C50 & $01\,08\,33.5$ & $+33\,47\,10$ & $123.81$ & $4$ \\
C51 & $01\,09\,40.6$ & $+34\,14\,13$ & $120.46$ & $4$ \\
C52 & $01\,10\,50.8$ & $+44\,44\,38$ & $84.72$ & $4$ \\
C55 & $01\,14\,29.6$ & $+46\,06\,07$ & $102.46$ & $4$ \\
C56 & $01\,17\,36.0$ & $+46\,11\,18$ & $109.10$ & $4$ \\
C58 & $01\,19\,43.8$ & $+33\,09\,20$ & $149.57$ & $4$ \\
C59 & $01\,22\,56.7$ & $+42\,14\,39$ & $103.27$ & $4$ \\
C60 & $01\,26\,10.8$ & $+43\,49\,11$ & $114.66$ & $4$ \\
C61 & $01\,27\,37.6$ & $+38\,07\,05$ & $125.50$ & $4$ \\
C63 & $01\,28\,38.6$ & $+44\,00\,47$ & $121.18$ & $4$ \\
C65 & $01\,31\,17.4$ & $+45\,43\,43$ & $134.68$ & $4$ \\
C66 & $01\,32\,45.8$ & $+42\,57\,32$ & $128.74$ & $4$ \\
C67 & $01\,33\,59.1$ & $+42\,38\,02$ & $131.41$ & $4$ \\
C68 & $01\,34\,06.0$ & $+45\,43\,32$ & $140.88$ & $4$ \\
C69 & $01\,34\,39.5$ & $+44\,05\,41$ & $135.82$ & $4$ \\
\hline
\label{t:zinn2}
\end{tabular}
\end{table}
 
During the preparation of this paper, \citet[][hereafter dTZZ13]{dtz:13} released a catalogue of M31 GCs and GC candidates derived from SDSS imaging. The area covered by SDSS overlaps substantially with the PAndAS footprint, allowing us to check the identity of many of the objects in the dTZZ13 catalogue -- although a number also lie well beyond the edge of the PAndAS coverage. 

The dTZZ13 catalogue consists of two primary lists. The first contains $18$ objects classified as high confidence GCs, while the second contains $75$ lower confidence candidate GCs. We located $17$ of the high confidence targets in our PAndAS imaging, along with $42$ of the candidates, and assessed these in the same manner as for objects in the RBC. The remaining dTZZ13 targets are at large radii from M31, $150 \la R_{\rm proj} \la 230$ kpc, and thus do not lie within the PAndAS footprint.
 
We found that ten of the $17$ high confidence objects that we inspected are indeed GCs, the remaining seven being either stars or distant galaxies. Classifications for these objects are listed in Table \ref{t:zinn1}. All but three of the ten GCs appear independently in our PAndAS catalogue, as indicated in the Table. The three outstanding clusters are at relatively
small projected radii, $R_{\rm proj} \la 20$ kpc, and thus fall within the inner limiting radius of our uniform search area. This adds further weight to the
suggestion from our RBC work above that a thorough search for GCs in the inner M31 halo may be fruitful. 
We show $g$-band thumbnails of the three new SDSS clusters in Figure \ref{f:sdssGCs}.

Of the $42$ candidate objects inspected, we only confirmed one as a genuine GC. This is C62, which we also list in
Table \ref{t:zinn1} and which also appears in our PAndAS catalogue. All of the other candidate objects turned out to be background
galaxies; we list these in Table \ref{t:zinn2}. 

Based on a simple extrapolation of our results, it is moderately likely ($ \sim 60\%$) that the remaining high confidence object from
dTZZ13 -- SDSS17, which falls at $R_{\rm proj} \sim 158$ kpc -- is a GC. However the success rate from their lower confidence 
candidate list is substantially smaller ($\sim 2.4\%$), suggesting that in the group of $33$ such objects that we were unable to inspect 
there may be at most one or two genuine GCs. Nonetheless, with $R_{\rm proj} > 150$ kpc, it would be very worthwhile tracking these down.

\section{Cluster Photometry and Sizes}
\label{photometry}
\subsection{Integrated luminosities}
We performed aperture photometry on each of our $59$ newly-discovered GCs using the {\it phot} task in {\sc iraf}. We also photometered our two GC candidates, the additional $6$ newly-confirmed GCs from the RBC and dTZZ13 listed in Tables \ref{t:rbc1} and \ref{t:zinn1}, SH06, and, to ensure a complete uniform sample of measurements for the outer M31 system, all other known GCs lying at $R_{\rm proj} \geq 25$\ kpc ($38$ objects, predominantly from  Hux08). Our results may be found in Table \ref{t:phot}. 

For each target, we used {\it phot} to measure the flux in concentric apertures of increasing radius, and constructed a curve-of-growth.
We employed the centroiding algorithm in the {\it phot} task to accurately determine the cluster centres\footnote{Note that our calculated coordinates for PA-31, and the handful of other GCs detected by \citet{Martinetal13}, are somewhat different than those listed in that paper. This is because Martin et al. report the coordinates of the spatial grid point in their calculation corresponding to the local probability maximum. The coordinates determined here are more accurate.}. This worked very well except on the most 
diffuse objects in our sample, which are fully resolved in the PAndAS imaging. For such targets we determined the centroid by eye, and verified that
our photometric measurements were robust to changes of a few pixels ($\sim 0.5\arcsec$) in any direction about this point. For each GC
we combined the central coordinates determined (independently) from the $g$- and $i$-band images in a straight average, and these are the positions 
reported in Table \ref{tab:locations}. In all cases the difference in coordinates from the $g$- and $i$-band images was less than $0.3\arcsec$, and 
in most cases less than $0.1\arcsec$. To estimate the background flux for a given GC we used the median level in an annulus of width $10\arcsec$ sitting 
outside the selected maximum photometry aperture. In practice the precise position of this background annulus was necessarily determined 
iteratively together with the maximum aperture itself. 

\begin{figure*}
\begin{center}
\includegraphics[width=53mm]{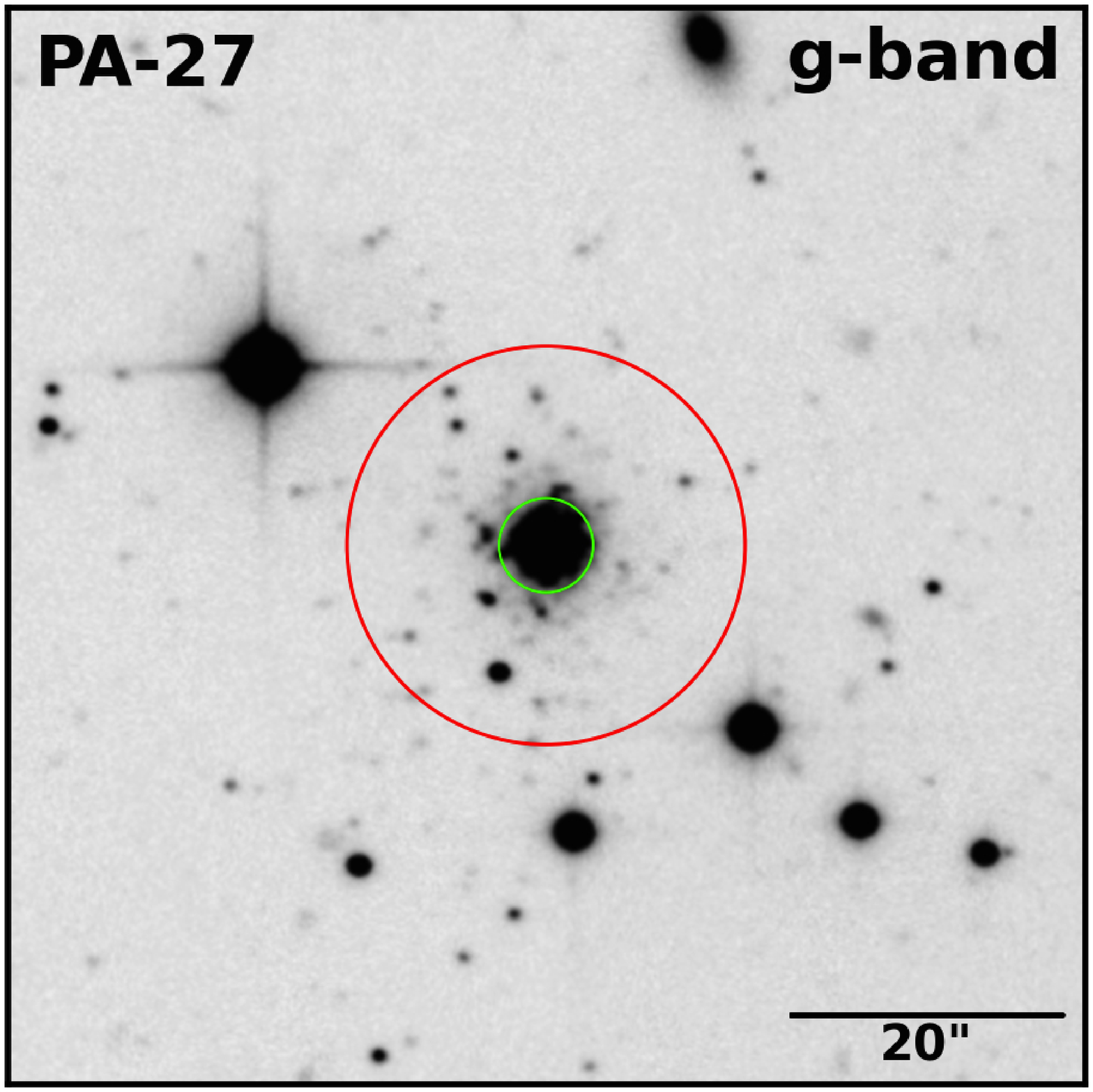}
\hspace{0mm}
\includegraphics[width=60mm]{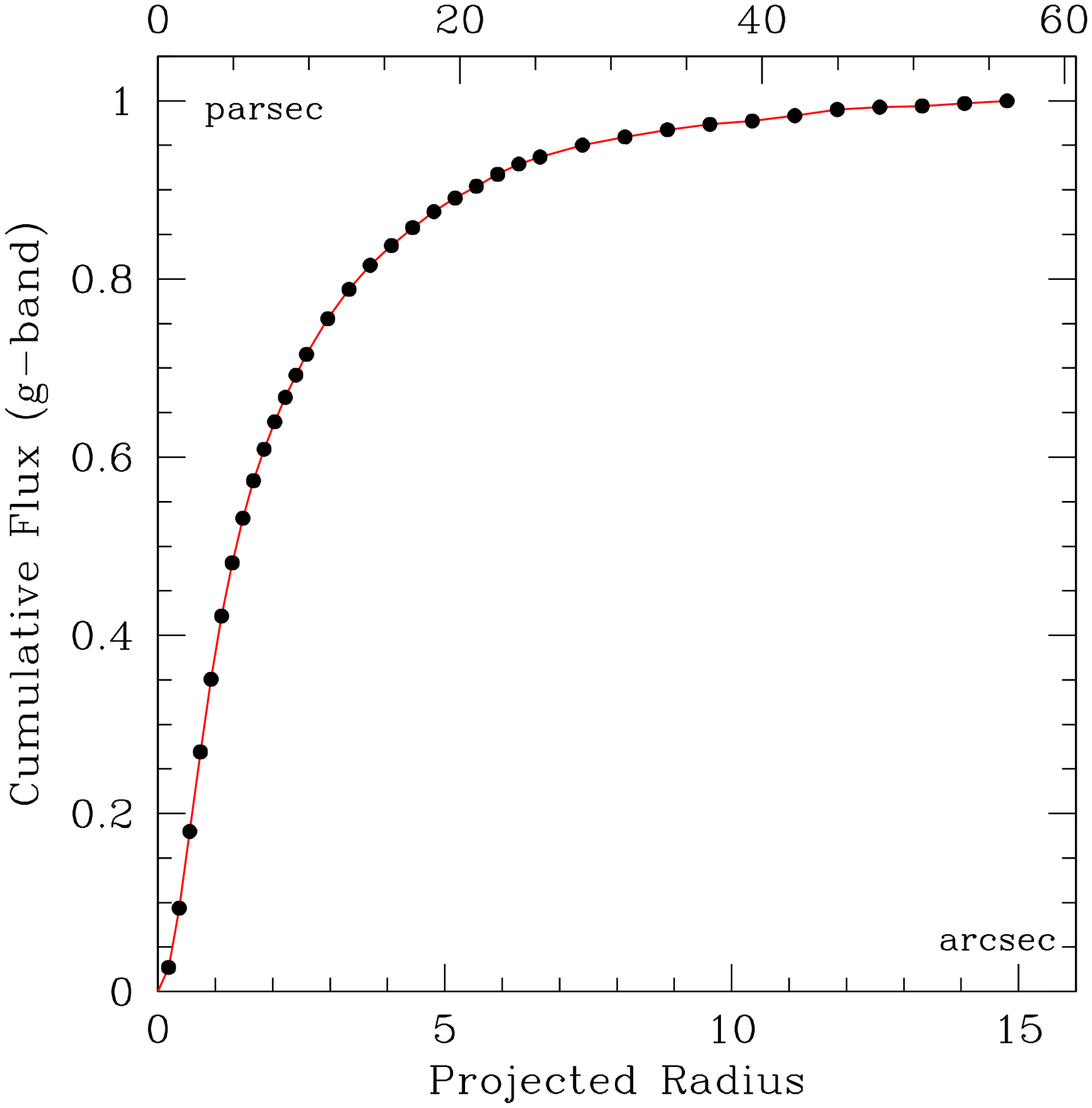}
\hspace{-1mm}
\includegraphics[width=60mm]{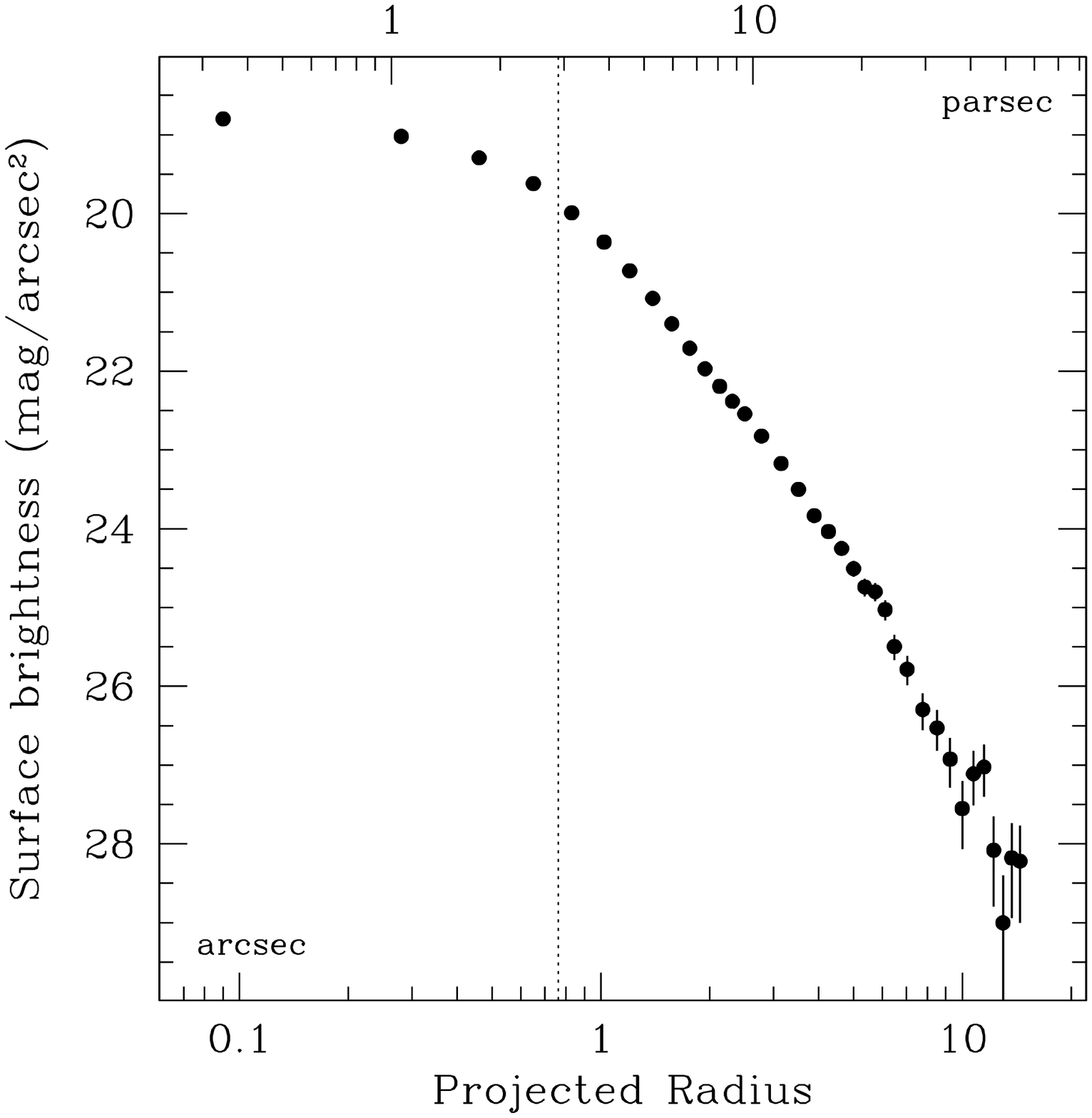}
\caption{Example of our photometric measurements for PAndAS-27. The left panel shows the $g$-band image of the cluster, with the maximum aperture ($r_{\rm max} = 14.8\arcsec$) marked in red, and the colour aperture ($3.5\arcsec$) marked in green. North is to the top of the page, and east to the left; the moderately bright foreground star within $r_{\rm max}$ to the SSE of the cluster was masked during the procedure. The PSF FWHM is slightly broader than the PAndAS $g$-band median at $0.76\arcsec$. The central panel shows the curve-of-growth for PA-27; note that this has clearly levelled out by the time the maximum aperture is reached. The measured half-light radius is $r_h = 1.4\arcsec  \sim 5.2$ pc. The right panel shows the curve-of-growth converted to a radial surface-brightness profile. The PSF FWHM is marked with a vertical dotted line; note that the profile flattens rather abruptly within this radius.}
\label{f:phot}
\end{center}
\end{figure*}

For an isolated cluster with little or no foreground or background contamination, we would define the maximum aperture $r_{\rm max}$ to sit
at a point where the increase in cumulative flux with radius (i.e., the curve-of-growth) is flat -- thus ensuring the inclusion of essentially all 
cluster light in the measurement. Figure \ref{f:phot} shows an example for the GC PA-27. However, only $\sim 60\%$ of our
systems conform to this ideal. A few objects are badly impacted by their proximity to the edge of a CCD (e.g., B517), a very bright 
star (H13, PA-9) or galaxy (HEC11), or, for more centrally located clusters ($R_{\rm proj} < 25$ kpc), the presence of moderately dense M31 field populations 
(e.g., PA-32, PA-35). Such cases cannot easily be corrected and thus for this type of object we were forced to artificially constrain $r_{\rm max}$ 
to a point on the curve-of-growth where the gradient is not necessarily flat, leading to an under-estimate in the flux. Note that wherever 
possible in this situation we kept the background annulus substantially outside the enforced limiting radius for photometry so as to avoid any 
cluster contribution to the estimated background level -- although in such cases this would never be the dominant source of uncertainty in 
any event. On a few rare occasions (e.g., PA-51, PA-55) a cluster fell so close to a CCD edge that it was only (partially) visible in one passband. 
In this situation useful photometry is not possible.

The most common non-ideal scenario we encountered fell between the two extremes of a completely isolated cluster and an object severely 
impacted by a chip edge or an excessively bright local contaminant. Typically, a given target might have an unrestricted maximum aperture, 
but a few ($\la 5$) sources lying within this aperture that were obviously either background galaxies or foreground stars of sufficient 
brightness to noticeably impact the measured flux. In general we found it straightforward to mask these objects such that the affected pixels 
were not used in the flux calculation. We also note that a few clusters (e.g., G1, PA-53) are sufficiently bright so as to be mildly saturated at 
their centres in the PAndAS imaging. While this affects the shape of the curve-of-growth at small radii, in no case was the saturation severe 
enough to alter the total flux measured within $r_{\rm max}$.

Given the variety of different circumstances seen across our sample, we assigned a flag to each object to indicate the quality of the photometric
measurement. These sit on a scale of A to D, with the following meanings:
\begin{itemize}
\item{A. An ideal isolated cluster, with an unrestricted maximum aperture and limited or no masking of contaminant sources necessary.}
\item{B. Minor issues, such as the necessity for moderate masking of contaminants, or a slightly restricted maximum aperture due to a CCD edge, 
nearby bright star, or non-trivial field background -- but not sufficient to under-estimate the flux by more than $\sim0.1-0.2$ mag.}
\item{C. Major issues and potentially significant unreliability, due to, for example, scattered light from a very nearby bright star or galaxy,
a strongly limited maximum aperture, or a contaminant coincident with the cluster centre, the masking of which interfered substantially
with the cluster flux.}
\item{D. Fatal problems, such as the majority of the cluster falling off the edge of a CCD, or the centre falling precisely coincident with a bright 
contaminant that could not be masked. Useful photometry is not possible for objects in this category.}
\end{itemize}
The quality flags are included in Table \ref{t:phot} along with notes indicating the specific issues, if any, arising for each particular GC
(for example, whether $r_{\rm max}$ was truncated, and if so why). For any analysis utilising our photometric measurements, only objects in 
categories A and B should be used. Photometry for objects flagged with a C is useful only for determining indicative properties such as 
whether a cluster is ``bright'' or ``faint'', or ``compact'' or ``diffuse''. 

Because all the GCs for which we derived photometry are either brand new, or sit at large galactocentric distances, there is minimal overlap between 
our sample and the set of objects possessing high precision luminosity measurements in the literature.  We found eight compact category 
`A'  or `B' clusters in our sample for which luminosities were measured from {\it HST} imaging by \citet{Tanviretal12} -- H1, H4, H5, H10, 
H23, H24, H27, and B514. Because the {\it HST} imaging is in different filters than our PAndAS data, we compare the integrated absolute 
$V$-band luminosities, $M_V$, calculated from the total $g$- and $i$-band magnitudes as detailed in Section \ref{ss:phottrans} below.
The mean offset in $M_{V}$ between our measurements and those from Tanvir et al. is $+0.09$ mag, and the dispersion about this value is
$0.13$ mag. Our luminosities are typically a little fainter than the {\it HST} measurements, which is not surprising as resolved photometry 
allows cluster members to be isolated even at radii well beyond our adopted $r_{\rm max}$ values.

We located four additional compact category `A'  or `B' clusters in our sample that have previous luminosity measurements from 
{\it HST} imaging calculated by \citet{Barmbyetal07} -- G1, G2, G339, and G353. When added to the Tanvir et al. clusters, the mean 
offset in $M_{V}$ between our measurements and those from the literature drops to $+0.01$ mag, but the dispersion rises somewhat
to $0.18$ mag.

Finally, there are two very diffuse clusters in our sample that were measured by Tanvir et al. -- HEC7 and HEC12. For these two objects we find
$M_{V}$ to be more substantially under-estimated, by $0.46$ and $0.52$ mag respectively. It is not clear why our $M_{V}$ estimates are 
$\sim 0.5$ mag fainter than the {\it HST} values -- most likely this reflects an inherent systematic limitation in integrating the extremely faint 
diffuse light component of these objects on medium-deep ground-based imaging.
 
\subsection{Size estimates}
In addition to determining the GC luminosities, we also used the curves-of-growth to obtain an empirical measure of their
structures -- as parametrised by the half-light radius $r_h$, which is the projected radius of an aperture encircling half a cluster's flux. 
We report $r_h$ for each GC in Table \ref{t:phot}; this quantity for a given target is the straight average of the (independent) measurements from
the $g$- and $i$-band images\footnote{Although $r_h$ may, in principle, be intrinsically slightly variable between various passbands (if, for example, a GC
is mass segregated), in practice our individual measurement errors of $\ga 10\%$ (see text) are dominant. We take the mean of the two size estimates to
try and minimise this random uncertainty.}. The quoted sizes are not meant to represent extremely precise measurements of the cluster 
structures -- performing such work on distant objects such as these using ground-based imaging is challenging and complex, and 
beyond the scope of the present paper. Rather, our estimates of $r_h$ are intended to provide a quantitative indication of whether a GC
is compact or diffuse, or somewhere in between. It has been known for some time that the halo of M31 hosts numerous unusually
extended clusters \citep[see e.g.,][]{Huxoretal05,Huxoretal11}, and it is thus of interest to be able to examine, even if just at an indicative level, 
the distribution of GC sizes across the complete sample.

We first note that our method of estimating $r_h$ is robust only if $r_{\rm max}$ falls on the flat part of the curve-of-growth. If not, then 
both the total luminosity and the half-light radius will be under-estimated. As described above, clusters for which $r_{\rm max}$ was truncated 
are flagged in the table; quoted sizes for these objects should be treated with caution. 

An additional, and arguably more important factor to consider is the effect of the seeing profile on our size measurements. Compact GCs in both the Milky Way and M31 have $r_h  \sim 3-5$ pc \citep[e.g.,][]{Harrisetal96,Tanviretal12}, which corresponds to an 
angular size of $\sim 0.8-1.3 \arcsec$ at the M31 distance ($\mu = 24.47$). This is not much larger than the mean 
full-width at half-maximum (FWHM) of point sources in PAndAS imaging, which is $0.67 \arcsec$ in $g$ and
$0.60 \arcsec$ in $i$. Thus observations of $r_h$ for very compact clusters in our sample largely reflect the seeing profile of the
PAndAS image in which the object falls, rather than the intrinsic properties of the GC. In principle this problem may be corrected by
careful deconvolution of the local image point-spread function (PSF) and the radial brightness profile of the cluster; this problem will be addressed 
by an upcoming analysis of a substantial new {\it HST} dataset (Mackey et al. 2014, in prep.). For now, we used artificial GC images
generated to assess the completeness of our PAndAS catalogue (see Section \ref{completeness}) to explore the impact of the PSF on our size measurements.

\begin{figure}
\begin{center}
\includegraphics[width=80mm]{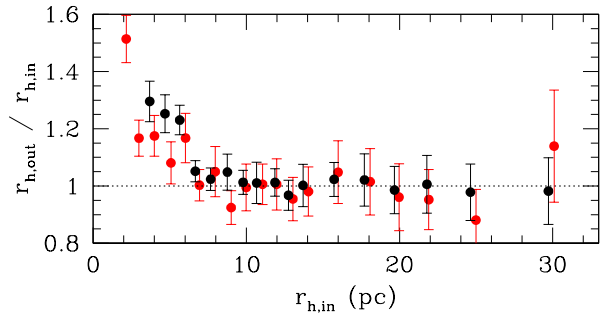}
\caption{Measured versus input half-light radii for artificial clusters with luminosities $M_{V} \sim -8$ (black points) and $\sim -5.5$ (red points).
For clarity the black points have been offset slightly along the $x$-axis.}
\label{f:rhfake}
\end{center}
\end{figure}

\begin{figure}
\begin{center}
\includegraphics[width=80mm]{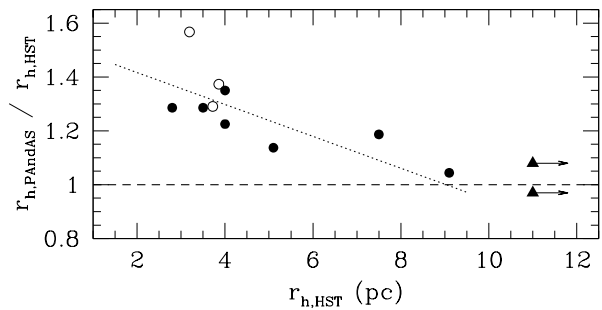}
\caption{Half-light radii derived in this paper versus those derived from {\it HST} imaging by \citet{Barmbyetal07} (open points) and \citet{Tanviretal12} (filled points).  Circles represent compact GCs, while triangles are diffuse GCs. Note that the two triangles should sit at $\sim 20$ pc and $\sim 30$ pc, but have been plotted at smaller radii to maintain clarity. The inclined dotted line represents a straight linear fit to all points in Figure \ref{f:rhfake} for which the input size was below $9$ pc. The apparently deviant point from \citet{Barmbyetal07} is G2, which is mildly saturated in our images (pushing $r_h$ to a larger value). The point for G1 falls well off the top of the plot as it is quite strongly saturated in the PAndAS images.}
\label{f:rhlit}
\end{center}
\end{figure}

\begin{table*}
\centering
\caption{Photometric measurements for PAndAS globular clusters and selected others.}
\begin{tabular}{@{}lcccccccccl}
\hline \hline
Cluster & $E(B-V)$ & $r_{\rm max}$ & $g$ & $i$ & $(g-i)_0$ & $M_{V}$ & $(V-I)_0$ & $r_h$ & Quality & Notes \\
Name & & (arcsec) & & & & & & (pc) & Flag & \\
\hline
PAndAS-1 & 0.099 & 14.8 & 17.55 & 16.74 & 0.62 & -7.48 & 0.83 & 7.1 & A & $...$ \\
PAndAS-2 & 0.106 & 19.2 & 18.30 & 17.29 & 0.70 & -6.82 & 0.90 & 25.7 & A & $...$ \\
PAndAS-3 & 0.087 & 13.9 & 20.89 & 19.90 & 0.65 & -4.17 & 0.86 & 27.4 & B & c \\
PAndAS-4 & 0.133 & 14.1 & 18.07 & 17.17 & 0.69 & -7.09 & 0.89 & 4.7 & A & $...$ \\
PAndAS-5 & 0.078 & 11.8 & 19.94 & 19.08 & 0.79 & -5.05 & 0.97 & 17.0 & A & c \\
PAndAS-6 & 0.068 & 14.1 & 16.92 & 16.15 & 0.67 & -8.02 & 0.87 & 4.4 & A & $...$ \\
PAndAS-7 & 0.088 & 11.8 & 20.18 & 18.77 & 0.70 & -5.00 & 0.89 & 13.3 & A & c \\
PAndAS-8 & 0.109 & 10.4 & 19.89 & 18.29 & 0.87 & -5.40 & 1.03 & 9.5 & A & $...$ \\
PAndAS-9 & 0.090 & 5.2 & 18.23 & 17.51 & 0.62 & -6.75 & 0.83 & 3.8 & C & b,r$^*$ \\
PAndAS-10 & 0.094 & 12.2 & 19.61 & 18.75 & 0.75 & -5.43 & 0.93 & 15.2 & A & c \\
PAndAS-11 & 0.088 & 12.6 & 18.29 & 17.41 & 0.67 & -6.74 & 0.87 & 9.4 & A & $...$ \\
PAndAS-12 & 0.060 & 7.4 & 19.63 & 18.72 & 0.75 & -5.33 & 0.94 & 13.7 & B & e,r \\
PAndAS-13 & 0.063 & 7.4 & 18.43 & 17.66 & 0.65 & -6.49 & 0.85 & 4.7 & B & m,b,r \\
PAndAS-14 & 0.069 & 12.6 & 17.93 & 17.17 & 0.71 & -7.01 & 0.90 & 10.9 & A & $...$ \\
PAndAS-15 & 0.069 & 3.7 & 19.92 & 19.08 & 0.74 & -5.04 & 0.93 & 6.0 & C & b,e,r$^*$ \\
PAndAS-16 & 0.072 & 16.3 & 16.54 & 15.66 & 0.79 & -8.44 & 0.97 & 5.5 & A & $...$ \\
PAndAS-17 & 0.067 & 14.1 & 16.87 & 15.77 & 1.00 & -8.17 & 1.14 & 4.4 & A & $...$ \\
PAndAS-18 & 0.062 & 14.8 & 19.61 & 18.71 & 0.76 & -5.35 & 0.94 & 23.0 & A & c \\
PAndAS-19 & 0.055 & 8.1 & 20.17 & 19.39 & 0.72 & -4.73 & 0.91 & 7.3 & A & $...$ \\
PAndAS-20 & 0.067 & 8.1 & 19.57 & 18.58 & 0.83 & -5.43 & 1.00 & 7.4 & A & $...$ \\
PAndAS-21 & 0.054 & 14.1 & 17.84 & 17.06 & 0.67 & -7.06 & 0.87 & 4.0 & A & $...$ \\
PAndAS-22 & 0.063 & 10.4 & 18.79 & 17.87 & 0.91 & -6.18 & 1.06 & 7.3 & A & $...$ \\
PAndAS-23 & 0.054 & 6.7 & 19.98 & 18.89 & 1.04 & -5.02 & 1.17 & 6.7 & A & $...$ \\
PAndAS-24 & 0.059 & 11.8 & 20.27 & 19.39 & 0.73 & -4.68 & 0.91 & 16.9 & A & c \\
PAndAS-25 & 0.064 & 4.8 & 19.78 & 18.80 & 0.88 & -5.21 & 1.04 & 6.6 & B & b,r \\
PAndAS-26 & 0.061 & 5.6 & 19.88 & 18.92 & 0.94 & -5.10 & 1.09 & 7.4 & B & e,r \\
PAndAS-27 & 0.075 & 14.8 & 17.31 & 16.41 & 0.75 & -7.69 & 0.93 & 5.2 & A & $...$ \\
PAndAS-28 & 0.066 & 8.3 & 19.26 & 18.56 & 0.64 & -5.65 & 0.85 & 12.7 & B & f,r \\
PAndAS-29 & 0.058 & 5.2 & 20.58 & 19.75 & 0.72 & -4.35 & 0.91 & 10.7 & B & c,f,r \\
PAndAS-30 & 0.064 & 7.4 & 19.57 & 18.58 & 0.79 & -5.42 & 0.96 & 10.9 & A & c \\
PAndAS-31 & 0.073 & 9.3 & 20.62 & 19.60 & 0.82 & -4.41 & 0.99 & 18.5 & B & c,m \\
PAndAS-32 & 0.075 & 6.7 & 19.48 & 18.37 & 0.97 & -5.58 & 1.11 & 8.0 & B & f,r \\
PAndAS-33 & 0.059 & 18.5 & 19.56 & 18.67 & 0.68 & -5.39 & 0.88 & 35.8 & B & c,b \\
PAndAS-34 & 0.068 & 12.6 & 18.37 & 17.38 & 0.81 & -6.64 & 0.98 & 10.4 & A & $...$ \\
PAndAS-35 & 0.086 & 5.2 & 19.91 & 18.59 & 1.09 & -5.24 & 1.21 & 10.3 & B & c,f,r \\
PAndAS-36 & 0.073 & 10.4 & 17.69 & 16.79 & 0.75 & -7.30 & 0.94 & 5.8 & A & $...$ \\
PAndAS-37 & 0.057 & 9.6 & 17.66 & 16.56 & 1.00 & -7.35 & 1.13 & 4.6 & A & $...$ \\
PAndAS-38 & 0.159 & 13.9 & 20.76 & 19.75 & 0.61 & -4.50 & 0.83 & 24.4 & B & c,m \\
PAndAS-39 & 0.086 & 9.6 & 18.85 & 17.90 & 0.88 & -6.19 & 1.04 & 13.0 & B & c,f,r \\
PAndAS-40 & 0.058 & 10.0 & 19.80 & 18.97 & 0.77 & -5.13 & 0.95 & 10.0 & A & $...$ \\
PAndAS-41 & 0.096 & $-$ & $-$ & $-$ & $-$ & $-$ & $-$ & $-$ & D & e,r$^*$ \\
PAndAS-42 & 0.060 & 15.5 & 18.54 & 17.04 & 0.89 & -6.59 & 1.05 & 15.4 & C & c,b \\
PAndAS-43 & 0.093 & 4.4 & 19.79 & 18.85 & 0.79 & -5.27 & 0.97 & 6.0 & B & c,e,r \\
PAndAS-44 & 0.062 & 9.6 & 17.18 & 16.48 & 0.61 & -7.72 & 0.82 & 3.1 & A & $...$ \\
PAndAS-45 & 0.083 & 7.4 & 20.97 & 20.05 & 0.79 & -4.06 & 0.96 & 8.9 & B & c \\
PAndAS-46 & 0.072 & 16.3 & 16.27 & 15.52 & 0.64 & -8.67 & 0.85 & 4.3 & B & s \\
PAndAS-47 & 0.070 & 5.6 & 19.39 & 18.26 & 1.01 & -5.66 & 1.14 & 3.8 & A & $...$ \\
PAndAS-48 & 0.066 & 13.7 & 20.21 & 19.41 & 0.59 & -4.73 & 0.81 & 21.2 & A & c \\
PAndAS-49 & 0.067 & 11.1 & 20.24 & 19.11 & 0.92 & -4.81 & 1.07 & 16.4 & A & $...$ \\
PAndAS-50 & 0.163 & 14.8 & 18.93 & 17.78 & 0.95 & -6.38 & 1.10 & 17.1 & A & $...$ \\
PAndAS-51 & 0.074 & $-$ & $-$ & $-$ & $-$ & $-$ & $-$ & $-$ & D & e,x \\
PAndAS-52 & 0.063 & 13.3 & 17.38 & 16.49 & 0.78 & -7.58 & 0.96 & 6.5 & B & b \\
PAndAS-53 & 0.053 & 12.6 & 15.79 & 15.07 & 0.64 & -9.09 & 0.85 & 4.2 & B & s,b,r \\
PAndAS-54 & 0.053 & 12.6 & 16.30 & 15.57 & 0.63 & -8.58 & 0.84 & 5.1 & C & b,m \\
PAndAS-55 & 0.070 & $-$ & $-$ & $-$ & $-$ & $-$ & $-$ & $-$ & D & e,x \\
PAndAS-56 & 0.050 & 14.1 & 17.27 & 16.45 & 0.70 & -7.63 & 0.89 & 4.7 & A & $...$ \\
PAndAS-57 & 0.066 & 8.9 & 19.24 & 18.44 & 0.71 & -5.70 & 0.91 & 10.3 & A & c \\
PAndAS-58 & 0.062 & 11.5 & 18.82 & 17.83 & 0.88 & -6.17 & 1.04 & 9.3 & A & $...$ \\
PAndAS-59 & 0.068 & 3.7 & 19.96 & 19.32 & 0.53 & -4.93 & 0.76 & 5.6 & B & f,r \\
PAndAS-Ca1 & 0.067 & 4.4 & 20.84 & 20.27 & 0.45 & -4.03 & 0.70 & 8.1 & B & f,r \\
PAndAS-Ca2 & 0.175 & 4.4 & 20.02 & 19.08 & 0.65 & -5.26 & 0.86 & 10.4 & C & e,r$^*$ \\
\hline
\label{t:phot}
\end{tabular}
\end{table*}

\addtocounter{table}{-1}
\begin{table*}
\centering
\caption{Continued.}
\begin{minipage}{140mm}
\begin{tabular}{@{}lcccccccccl}
\hline \hline
Cluster & $E(B-V)$ & $r_{\rm max}$ & $g$ & $i$ & $(g-i)_0$ & $M_{V}$ & $(V-I)_0$ & $r_h$ & Quality & Notes \\
Name & & (arcsec) & & & & & & (pc) & Flag & \\
\hline
G1 & 0.057 & 32.6 & 14.17 & 13.23 & 0.77 & -10.79 & 0.95 & 8.7 & B & s,b,m \\
G2 & 0.052 & 25.2 & 15.97 & 15.21 & 0.67 & -8.92 & 0.87 & 5.0 & B & s \\
G339 & 0.093 & 11.8 & 17.47 & 16.55 & 0.75 & -7.58 & 0.94 & 5.3 & A & $...$ \\
G353 & 0.082 & 10.4 & 17.39 & 16.58 & 0.67 & -7.60 & 0.87 & 4.8 & A & $...$ \\
B514 & 0.058 & 25.2 & 16.02 & 15.17 & 0.77 & -8.91 & 0.95 & 6.6 & A & $...$ \\
B517 & 0.065 & $-$ & $-$ & $-$ & $-$ & $-$ & $-$ & $-$ & D & e,r$^*$ \\
EXT8 & 0.068 & 15.5 & 15.79 & 14.58 & 0.56 & -9.28 & 0.79 & 4.3 & B & s \\
MGC1 & 0.086 & 37.4 & 15.60 & 14.15 & 0.71 & -9.59 & 0.91 & 8.8 & A & $...$ \\
H1 & 0.070 & 18.5 & 16.25 & 15.45 & 0.68 & -8.70 & 0.88 & 4.5 & B & s \\
H2 & 0.059 & 18.5 & 17.43 & 16.60 & 0.73 & -7.50 & 0.91 & 5.2 & A & $...$ \\
H3 & 0.069 & 8.9 & 18.48 & 17.53 & 0.85 & -6.52 & 1.01 & 5.5 & A & $...$ \\
H4 & 0.073 & 17.0 & 17.17 & 16.28 & 0.74 & -7.82 & 0.93 & 5.4 & A & $...$ \\
H5 & 0.075 & 17.0 & 16.51 & 15.76 & 0.64 & -8.44 & 0.85 & 9.5 & A & $...$ \\
H7 & 0.057 & 17.0 & 17.76 & 16.92 & 0.73 & -7.17 & 0.92 & 10.5 & A & $...$ \\
H8 & 0.062 & 11.1 & 19.33 & 18.18 & 0.87 & -5.71 & 1.03 & 11.8 & B & f \\
H9 & 0.055 & $-$ & $-$ & $-$ & $-$ & $-$ & $-$ & $-$ & D & e,x \\
H10 & 0.065 & 23.7 & 16.25 & 14.88 & 0.77 & -8.86 & 0.95 & 5.8 & A & $...$ \\
H11 & 0.071 & 11.8 & 17.10 & 16.23 & 0.75 & -7.88 & 0.93 & 4.1 & A & $...$ \\
H12 & 0.066 & 13.3 & 16.75 & 15.95 & 0.68 & -8.19 & 0.88 & 4.0 & A & $...$ \\
H15 & 0.057 & 11.8 & 18.46 & 17.20 & 0.62 & -6.60 & 0.83 & 10.3 & A & c \\
H17 & 0.052 & 9.6 & 17.85 & 16.47 & 0.79 & -7.23 & 0.96 & 3.3 & A & $...$ \\
H18 & 0.087 & 14.8 & 16.92 & 16.08 & 0.70 & -8.09 & 0.90 & 4.1 & A & $...$ \\
H19 & 0.059 & 7.4 & 17.66 & 16.77 & 0.74 & -7.29 & 0.92 & 4.9 & A & $...$ \\
H22 & 0.050 & 11.1 & 17.26 & 16.43 & 0.74 & -7.65 & 0.93 & 4.2 & B & g \\
H23 & 0.051 & 8.9 & 17.00 & 15.55 & 0.85 & -8.09 & 1.01 & 3.6 & B & b,r \\
H24 & 0.098 & 14.8 & 17.97 & 17.04 & 0.75 & -7.10 & 0.94 & 8.9 & A & $...$ \\
H25 & 0.094 & 14.8 & 17.12 & 16.21 & 0.76 & -7.93 & 0.94 & 5.8 & A & $...$ \\
H26 & 0.053 & 14.8 & 17.66 & 16.34 & 0.70 & -7.40 & 0.90 & 5.6 & A & $...$ \\
H27 & 0.055 & 14.8 & 16.66 & 15.39 & 0.66 & -8.39 & 0.86 & 4.9 & A & $...$ \\
HEC1 & 0.060 & 12.6 & 19.06 & 18.39 & 0.65 & -5.82 & 0.86 & 15.7 & A & $...$ \\
HEC2 & 0.055 & 12.2 & 19.51 & 18.03 & 0.78 & -5.60 & 0.96 & 12.4 & B & c,e,r \\
HEC3 & 0.070 & 15.5 & 19.63 & 18.71 & 0.83 & -5.36 & 1.00 & 17.6 & A & c \\
HEC6 & 0.073 & 18.5 & 19.09 & 18.12 & 0.75 & -5.92 & 0.94 & 26.7 & A & c \\
HEC7 & 0.087 & 16.7 & 18.48 & 17.53 & 0.82 & -6.57 & 0.99 & 19.5 & A & c \\
HEC10 & 0.106 & 20.4 & 18.97 & 17.98 & 0.79 & -6.14 & 0.97 & 22.5 & A & c \\
HEC11 & 0.048 & 8.9 & 18.41 & 17.03 & 0.70 & -6.65 & 0.90 & 14.6 & B & c,g,r \\
HEC12 & 0.049 & 20.4 & 18.93 & 17.48 & 0.82 & -6.16 & 0.99 & 29.9 & A & c \\
HEC13 & 0.048 & 13.7 & 19.48 & 18.27 & 0.64 & -5.54 & 0.85 & 20.7 & B & c,e,r \\
B270D & 0.087 & 6.7 & 17.77 & 16.86 & 0.76 & -7.26 & 0.94 & 6.6 & B & f,r \\
SK213C & 0.121 & 3.3 & 19.43 & 18.43 & 0.81 & -5.72 & 0.98 & 4.4 & B & f,r \\
SK255B & 0.070 & 8.1 & 18.01 & 17.00 & 0.89 & -7.01 & 1.04 & 5.6 & B & b,f,r \\
SH06 & 0.168 & 8.1 & 16.43 & 16.55 & -0.49 & -8.45 & 0.00 & 9.9 & A & $...$ \\
SDSS1 & 0.068 & 8.1 & 18.33 & 17.22 & 0.95 & -6.71 & 1.10 & 12.0 & B & m \\
SDSS3 & 0.066 & 6.7 & 19.06 & 17.95 & 1.03 & -5.98 & 1.16 & 7.2 & A & $...$ \\
SDSS6 & 0.079 & $-$ & $-$ & $-$ & $-$ & $-$ & $-$ & $-$ & D & e,x \\
\hline
\end{tabular}
\medskip
\vspace{-2mm}
\\
Notes: b=nearby bright star, poorly masked or not maskable; c=centroided by eye; e=affected by CCD edge; f=high field star density; g=nearby bright galaxy, poorly masked or not maskable; m=masking required for many contaminanting sources; r=restricted maximum aperture; r$^*$=severely restricted maximum aperture; s=saturated at centre; x=missing in one or both filters.\\
\end{minipage}
\end{table*}
 
We constructed two representative samples from the full suite of $4760$ artificial GCs. As we describe in detail in Section \ref{completeness}, 
these objects were generated by
first specifying a structure and luminosity, and then constructing a realistic image assuming the median PAndAS seeing. Both of our samples
contained GCs spanning the full range of input half-light radii $r_h  \sim 3-35$ pc, but for one ensemble the cluster luminosities
fell within the range $M_{V} = -8.0 \pm 0.3$ and for the other within $M_{V} = -5.5 \pm 0.3$. We passed each artificial cluster in these two
samples through our photometry pipeline. Note that we only selected objects that would have been classified in category `A' in terms of 
the quality of the photometric measurement.

The results of this process are presented in Figure \ref{f:rhfake}. The marked error-bar for a given size bin corresponds 
to the standard deviation in the measured GC sizes within that bin. It is clear that for GCs with input $r_h$ larger than $\sim 8-10$ pc,
we recover a very reasonable estimate of the object's size. This appears to be true irrespective of luminosity, although not surprisingly the 
scatter noticeably increases for lower-luminosity GCs compared to higher-luminosity GCs. Our tests indicate that typical uncertainties in the 
measured values of $r_h$ are $\la 8\%$ for $M_{V}  \sim -8$ objects and $\la 12\%$ for $M_{V}  \sim -5.5$ objects. These uncertainties 
increase to $\sim 20\%$ for low luminosity objects with very large $r_h$. 

Below $r_h \sim 8-10$ pc it is clear that the size measurements are significantly affected by the seeing profile, as expected. 
This limit corresponds to roughly three times the FWHM of the PSF used when constructing the artificial cluster images. It is interesting 
to note that while the measured $r_h$ values become increasingly different from the input values when moving to smaller sizes, 
within the limitations of the scatter the ordering is preserved. That is, a GC that is intrinsically more compact than another will still 
be measured as such by our photometry pipeline even when strongly affected by the PSF. We must bear in mind that the seeing 
profile does vary between PAndAS images, unlike for our artificial clusters; however, as previously reported the rms scatter about the 
mean seeing values is small ($ \sim 0.1\arcsec$). 

We return briefly to the sample of $12$ compact GCs and $2$ diffuse GCs for which high precision photometry and structural measurements
exist in the literature. Figure \ref{f:rhlit} shows a comparison between our $r_h$ measurements and those derived from {\it HST} imaging by 
\citet{Barmbyetal07} and \citet{Tanviretal12}.  This strongly resembles Figure \ref{f:rhfake} -- for cluster sizes below $\sim 8-10$ pc, our 
estimated $r_h$ values are clearly too large; however the correct ordering is preserved. Indeed our measurements appear to behave exactly
as predicted by the artificial cluster tests -- the inclined dotted line represents a straight linear fit to all points in Figure \ref{f:rhfake} with 
input size below $r_h = 9$ pc, and this provides an excellent description of how strongly our measured quantities deviate from those obtained 
via {\it HST}. As a final note, we see that for the two diffuse clusters our $r_h$ measurements match those from {\it HST} to better than 
$\sim 10\%$, consistent with our estimated uncertainties.

\subsection{Integrated colours}
Comparing our $g$- and $i$-band flux measurements for a given GC allows us to derive the integrated colour of that object.
In principle, we could calculate $(g-i)$ directly from the total luminosities measured within $r_{\rm max}$. However, previous work has 
found that employing a smaller aperture can lead to a more robust result \citep[e.g.,][]{Huxoretal09, Veljanoskietal13b}. This is perhaps not 
too surprising, as the larger the aperture for colour measurement, the more sensitive the result is to (i) the presence of unidentified 
contaminants, and (ii) the accuracy of the estimated background level. 

For the present work, there are some subtleties associated with determining an optimal colour aperture.
First, this methodology is predicated on the absence of intrinsic colour gradients within the target GCs. This appears reasonable
-- high resolution studies from {\it HST} have not revealed any such gradients \citep[e.g.,][]{Tanviretal12}. Second, our catalogue spans 
a very large range of cluster sizes, so it does not make sense to simply apply a uniform colour aperture across
the entire sample. A small aperture that might work well for a compact GC could lead to a very misleading result for a diffuse 
GC, as it would be extremely sensitive to the presence, or absence, of a handful of bright stars at the centre of such an object.
Much better is to define an aperture that samples the same region in every cluster -- for example, $r_h$; however this introduces a
third issue which is that, as we have already seen, compact GCs are strongly affected by the seeing profile. In the context of
a colour measurement, it is important to recognise that any difference in the PSF width between the $g$- and $i$-band images
leads to an artificial colour gradient at the centre of the object due to the differential redistribution of flux. If the colour aperture
is set to be too small, any such gradient will result in an erroneous measurement. 

Fortunately, we already know from Figure \ref{f:rhfake} the radius at which the effect of the seeing profile becomes negligible.
Thus, we set the colour aperture to be equal to $r_h$ for all GCs down to a conservative limit of $r_h = 3.5\arcsec  \sim 13$ pc; 
for any clusters with $r_h$ smaller than this, the colour aperture is set at $3.5\arcsec$. This lower limit matches the
uniform colour aperture used in comparable studies, such as that of \citet{ Veljanoskietal13b} for GCs in the M31 dwarf elliptical (dE)
companions NGC 147 and 185 (although note that that sample did not span anything like the range in size as does the present sample).

\subsection{Photometric transformations}
\label{ss:phottrans}
We conclude this section by summarising the measurements reported in Table \ref{t:phot}. For each GC we list the foreground colour 
excess $E(B-V)$ as derived from the \citet{Schlegeletal98} maps, and the maximum photometry aperture $r_{\rm max}$. Next, we list the total
integrated $g$- and $i$-band AB magnitudes within $r_{\rm max}$, along with the $(g-i)$ colour determined from a more central aperture 
as described above. The colour as reported has been dereddened using the appropriate coefficients from the study of \citet{Schlaflyetal11}:
\begin{eqnarray}
g_0 &=& g - 3.303\,E(B-V) \nonumber \\
i_0 &=& i - 1.698\,E(B-V)\,.
\end{eqnarray}
An important subtlety is that in mid-2007 the CFHT/MegaCam $i$-band filter was broken, and subsequently replaced with a new filter 
possessing a slightly different transmission profile. As a result, instrumental $i$-band magnitudes for GCs falling in images taken prior 
to June 2007 are calibrated to a slightly different system than for GCs taken after this date. To ensure a consistent set of 
measurements across the entire sample, we use the relationship from \citet{Ibataetal13} to transform the photometry for GCs imaged 
with the old $i$-band filter onto the system of the new filter. Since all our objects have $(g-i_{\rm old}) < 1.9$ this takes the form:
\begin{eqnarray}
i_{\rm new} &=& i_{\rm old} + 0.031\,(g-i_{\rm old}) - 0.010\,.
\label{e:ibata}
\end{eqnarray}
To facilitate comparison with GCs in the inner parts of M31, as well as in systems belonging to other galaxies, we also list in
Table \ref{t:phot} our photometry transformed to the standard Johnson-Cousins system using the relations from Hux08 
\citep[see also][]{ Veljanoskietal13b}. We first convert from AB magnitudes to Vega magnitudes:
\begin{eqnarray}
g_1 &=& g + 0.092 \nonumber \\
i_1 &=& i_{\rm old} - 0.401\,,
\label{e:ab2vega}
\end{eqnarray}
and then transform to $V$ and $I$:
\begin{eqnarray}
V &=& g_1 - 0.42\,(g_1 - i_1) + 0.04\,(g_1 - i_1)^2 + 0.10 \nonumber \\
I &=& i_1 - 0.08\,(g_1 - i_1) + 0.06\,.
\end{eqnarray}
Note that, as explicitly denoted in Equation \ref{e:ab2vega}, these relations are valid only for photometry in the {\it old} $i$-band 
system. Thus we transform all our measurements into this system using the inverse of Equation \ref{e:ibata} {\it prior} to implementing 
the above procedure. In Table \ref{t:phot} we list the absolute $V$-band magnitude $M_{V}$ and the dereddened $(V-I)$ colour. We calculate these
from $V$ and $I$ assuming a distance modulus $\mu = 24.47$, the relevant $E(B-V)$, and, as before, the appropriate coefficients
from \citet{Schlaflyetal11}:
\begin{eqnarray}
V_0 &=& V - 2.742\,E(B-V) \nonumber \\
I_0 &=& I - 1.505\,E(B-V)\,.
\end{eqnarray}

\section{Survey Completeness}
\label{completeness}
A thorough, quantitative assessment of detection completeness is critical to the utility of our globular cluster catalogue. We have identified 
two major sources of incompleteness associated with the PAndAS data and our search technique, and we quantify each of these below.

\subsection{Incomplete spatial coverage}
Although PAndAS does an excellent job of achieving uniform imaging of the M31 halo to $R_{\rm proj} \approx 120-150$\ kpc in all 
directions, there are myriad small gaps in its spatial coverage. These arise from two sources: (i) spaces between the first and second, 
and third and fourth rows of CCDs on the MegaCam focal plane, which were typically not filled in by the small telescope dithers employed 
during the observations; and (ii) imperfect tiling of the PAndAS mosaic.

In general this spatial incompleteness is of no consequence for the primary goal of the PAndAS survey -- studying the properties of the resolved 
M31 field halo and the large-scale substructures and overdensities that are found within it. Individual satellite dwarf galaxies of M31 are 
also typically large enough to span the missing regions. GCs, on the other hand, are sufficiently small on the sky that they can 
easily fall into a gap in the coverage and not be detected. We are aware of at least one such case of this occurring -- the object H9 from 
 Hux08, which sits at $R_{\rm proj} = 56$ kpc and was originally discovered in INT/WFC observations, does 
not appear in any PAndAS image because it sits squarely in an inter-chip gap.

Fortunately, this kind of incompleteness is straightforward to quantify. It affects all GCs equally, irrespective of their
morphology or luminosity; all that is required is to calculate the fraction of missing coverage as a function of projected galactocentric radius.
To do this, for every PAndAS image we used the WCS information in the header to determine the coordinates of the four corners of each of
the $36$ CCDs and hence the equations defining their edges in RA and Dec. Note that we treated $g$- and $i$-band images independently
in the list, as there are often small spatial offsets between images in the two filters and we only needed coverage from one filter to identify a GC.

Next, we constructed circular annuli about the M31 centre and 
filled each one with points generated at random positions so as to achieve uniform coverage of the area within the annulus. Each annulus
was of width $0.5$\ kpc at the M31 distance, or $0.0366\degr$ on the sky. We generated 
enough points per annulus to achieve a minimum density of $100\,$arcmin$^{-2}$ -- sufficient to properly sample the inter-row gaps between
CCDs ($ \sim 80\arcsec$ wide). Using the complete list of CCD edge equations, we tested each point to see whether it fell within the area
covered by any given chip, and hence within the imaged region of the PAndAS footprint. The fractional coverage for a particular annulus was
then simply the ratio of imaged to total points generated inside that annulus. 
 
\begin{figure}
\begin{center}
\includegraphics[width=80mm]{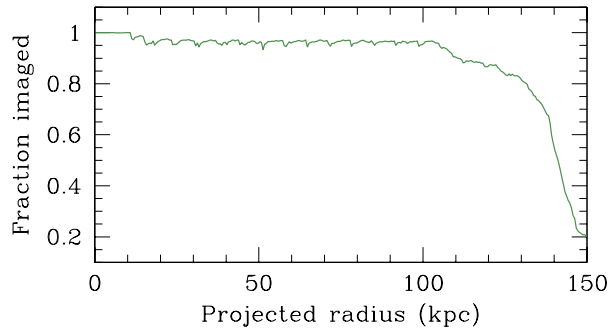}
\caption{Fractional spatial coverage of the PAndAS survey imaging as a function of projected radius from the M31 centre.}
\label{f:radialcompleteness}
\end{center}
\end{figure}

Figure \ref{f:radialcompleteness} shows our results. In the central regions of M31, within $R_{\rm proj}  \approx 10$\ kpc, there are sufficient 
overlapping images that the spatial coverage is complete. Beyond this, the coverage falls to $ \sim 96\%$ all the way out to
$R_{\rm proj} = 105$\ kpc, where the irregular edge of the PAndAS footprint gradually begins to affect the completeness. There is a shallow decline to 
$ \sim 80\%$ coverage at $R_{\rm proj} = 130$\ kpc, and then beyond this a rapid drop to $ \sim 20\%$ coverage at $150$\ kpc.

How does this affect our GC sample? Including previously-catalogued objects, we know of $82$ GCs with projected radii in the range
$25 \leq  R_{\rm proj} < 105$\ kpc; however only $81$ of these appear in PAndAS imaging (recall that H9 falls in an inter-chip gap). 
The $96\%$ spatial coverage over this radial range leads us to expect $84.4$ GCs, so we are likely missing just $2$ or $3$.
Given the radial decrease in the spatial density of GCs \citep[e.g.,][Mackey et al. 2014, in prep.]{ Huxoretal11}, and the uniform level of 
areal incompleteness, these missing objects are more likely to lie at smaller rather than larger projected radii. 

In the range $105 \leq  R_{\rm proj} < 130$\ kpc we know of $8$ GCs, all of which appear in the PAndAS imaging. A crude integration of the
completeness function suggests we are missing $ \sim 1$ additional GC in this range. Finally, in the range $130 \la  R_{\rm proj} < 150$\ kpc,
we have found only $1$ cluster in the PAndAS imaging \citep{Mackeyetal13a}; there is probably $\la 1$ other similarly remote object that falls outside the survey footprint.

To summarize, we have plausibly missed $\approx 3-5$ GCs over the range $25 \leq  R_{\rm proj} < 150$\ kpc due to the incomplete 
spatial coverage of the PAndAS imaging.

\subsection{Cluster identification/recognition}
The completeness of our catalogue is also affected by our ability to identify objects as GCs. That is,
it is certain that we miss some clusters due to them being too small, faint, compact, or diffuse (or some combination of these)
to recognise as GCs. There is also the possibility of human error to consider -- missing objects due to, say, a lapse in attention
while searching images.

All indications suggest that human error is a negligible factor for our search.
As a first pass, one of us (ADM) inspected $\sim 30\%$ of the images previously searched by APH, including a number with no 
GCs as well as some of those more heavily populated with GCs. In all cases the consistency of the results was excellent, suggesting
that our methodology, at least in uncrowded regions of low background, is robust. In addition to this, we recovered all known
GCs in our primary search area ($R_{\rm proj} \geq 25$\ kpc) -- from both the RBC \citep{galleti:04} and the previous INT survey (Hux08) 
-- with no omissions (barring H9). 
Finally, the automated search for dwarf spheroidal
satellites of M31 devised by NFM \citep{Martinetal13} recovered just one missed GC across our entire survey area. This object (PA-31) is diffuse and
very faint, falling near our $50\%$ completeness limit (see below) -- so its original omission is not surprising.
The search algorithm is sensitive only to objects possessing a sufficient number of resolved but relatively uncrowded stars. 
This describes just a
relatively small fraction of our final GC catalogue, but includes bright objects such as MGC1 \citep[e.g.,][]{Martinetal06,Mackeyetal10b},
as well as fainter extended clusters like PA-31. That it did not return a significant
number of missed systems is another indication that human error has not introduced appreciable incompleteness into our catalogue.

To quantify how our ability to identify GCs in PAndAS imaging is affected by cluster luminosity and structure, we used a sample of
artificial GCs. Ideally, these would be added into a wide variety of the PAndAS images themselves and then ``discovered'' (or not) via a
search methodology identical to that which we originally employed. This would have the added benefit of facilitating a more
precise quantification of any incompleteness arising due to human error, as well as that due to the presence of very bright foreground
stars (which we assume to be negligible due to the small number of such objects).
Unfortunately, however, this technique is not practical. To achieve barely viable statistics requires a minimum sample 
approaching $\sim 5000$ artificial clusters (see below). If distributed with comparable spatial density to the GCs in our
catalogue, the necessary search area would total $\sim 50$ times the area of the PAndAS footprint. Even distributing the artificial GCs
with an unrealistic factor of ten higher density would still require a search of several times the PAndAS footprint.

To circumvent this issue we employed a simpler technique. We constructed small thumbnails with our artificial 
GCs at the centre, one per thumbnail, and then inspected each of these with the aim of determining, as objectively as possible,
which would have been identified as a GC and which not. 
This methodology facilitated both our main aim of quantifying 
the faint limit of our survey, and our secondary aim of exploring how strongly this varies with cluster structure. 

Under the assumption that incompleteness 
due to human error is negligible, there ought to be no difference between results derived from our simple inspection technique and those 
derived via a full search for artificial GCs.  This assumption is a good approximation
for luminous and/or compact GCs -- objects which (i) were targeted by our inspection of colour-magnitude selected candidates, and (ii) 
were typically prominent and thus easily-located in the blind visual search.  However, the approximation may break down subtly for objects
of very low surface brightness because we knew, {\it a priori}, that each artificial thumbnail hosted an object at its centre.  This is in contrast to the 
real situation where it was necessary to first find these objects in the blind search\footnote{Recall that diffuse clusters typically did not
appear in the list of colour-magnitude selected candidates.}. Ultimately this mild systematic bias could mean that our derived faint
completeness limits are too generous by a few tenths of a magnitude -- and indeed, as we point out later, we may observe weak evidence
for such an effect.

Our artificial clusters were generated across a binned grid in luminosity and concentration\footnote{The concentration $c=\log(r_t / r_c)$ 
where $r_t$ is the cluster tidal radius and $r_c$ the core radius, assuming a \citet{king:62} model fit to the radial surface density
profile. A concentration $c=2.5$ typically indicates a GC affected by core collapse.}, extending between the limits 
$-10\leq M_{V} \leq -2$ and $0.5 \leq c \leq 2.5$. We generally adopted bin sizes of $0.25$ mag in $M_{V}$ and $0.1$ in concentration,
although for $M_{V} \leq -8$ and/or $c \geq 2.0$ we used bins of twice this size as finer discrimination was unnecessary. 
Within each bin we generated $10$ artificial clusters
with random $M_{V}$ and $c$, in order to uniformly sample the bin. This resulted in a total ensemble of $4760$ artificial GCs.
  
\begin{figure*}
\begin{center}
\includegraphics[width=150mm]{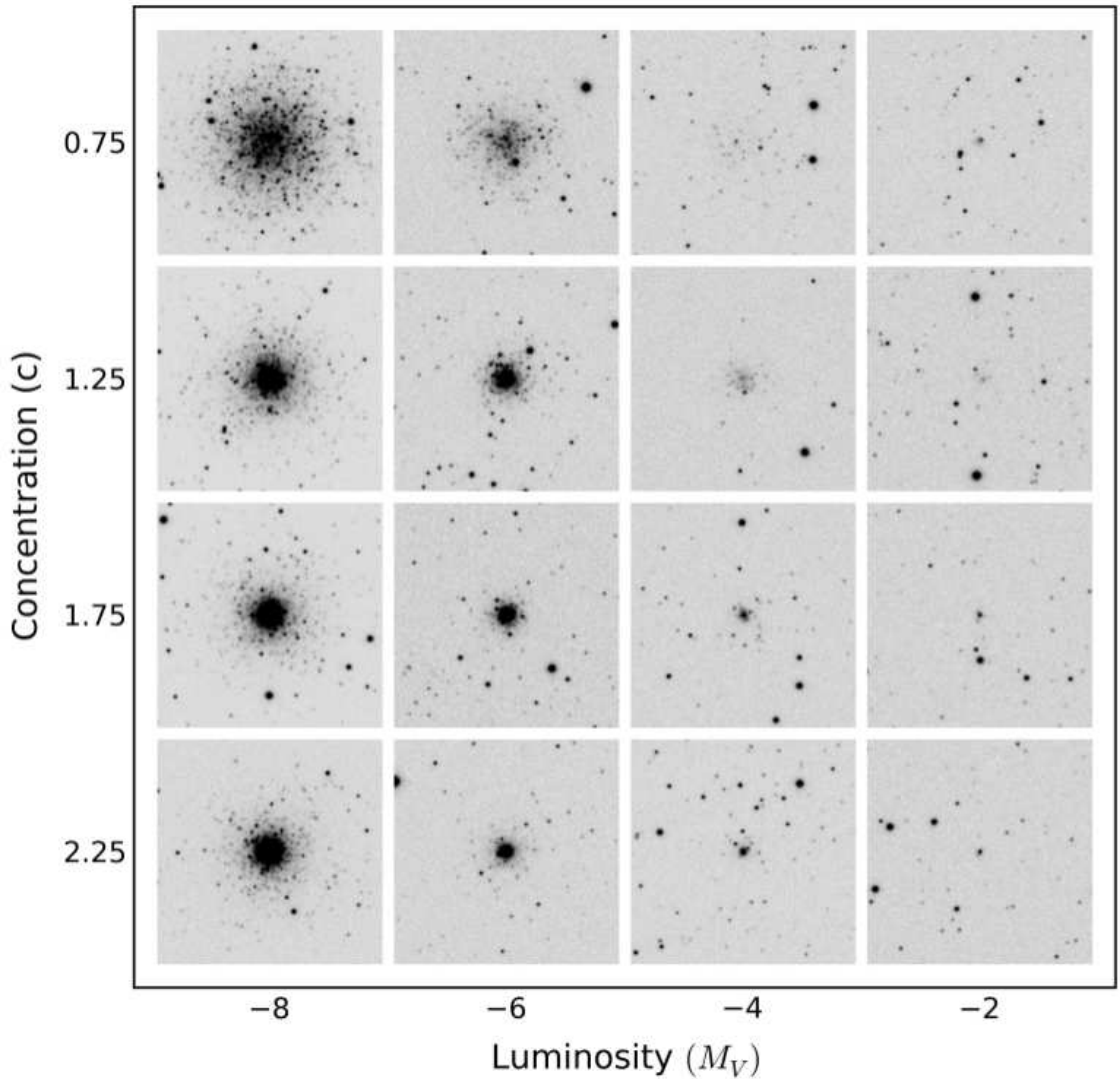}
\caption{Examples of $g$-band artificial cluster images across the luminosity-concentration plane. These are central $1\arcmin \times 1\arcmin$ cut-outs from our $1.6\arcmin \times 1.6\arcmin$  thumbnails. Each assumes a stellar FWHM of $0.7\arcsec$, typical for the vast majority of PAndAS $g$-band imaging.}
\label{f:fakeclusters}
\end{center}
\end{figure*}

We generated thumbnail images of these objects using the {\sc SimClust} software \citep{deveikis:08}. This package generates a random realisation
of a GC given its age, metallicity, mass ($M_{\rm cl}$), and structural parameters ($r_c$ and $r_t$), and then "observes" this model to produce
a realistic image. For simplicity we assumed a uniform age of $13$ Gyr and metallicity of $[$Fe$/$H$] = -1.8$ for all our artificial clusters. These 
values are representative of those observed for metal-poor halo GCs in both the Milky Way and M31 
\citep[e.g.,][]{Mackeyetal06,Mackeyetal07,marinfranch:09,dotter:10}. Cluster masses were set from the randomly generated luminosities
using $M/L = 2$, which is appropriate for the assumed age and metallicity. The structural parameters were determined by first assigning
to each GC a random 3D galactocentric radius within $25 \leq R_{\rm gc} \leq 145$\ kpc to match the range 
of projected radii observed for our PAndAS GC sample. This then defined the tidal radius according to the usual relationship 
$r_t = R_{\rm gc} (M_{\rm cl} / M_{\rm g})^{1/3}$ where we assumed a galactic mass $M_{\rm g} = 1.2\times 10^{12} M_\odot$ for M31,
and then our randomly generated concentration parameter determined $r_c$. Given this set of input parameters, {\sc SimClust} randomly selects 
stars from an appropriate Padova isochrone \citep{marigo:08} according to a set mass function, until the desired cluster mass is reached. We used 
the segmented power-law mass function of \citet{kroupa:01} for our GCs. The stars are randomly distributed spatially according to a \citet{king:62} 
model with appropriate $r_c$ and $r_t$. 

{\sc SimClust} converts all stellar positions and luminosities for a given artificial GC to ``observed'' quantities according 
to a specified distance and foreground extinction -- we used $\mu = 24.47$, and the typical colour excess across the PAndAS footprint of 
$E(B-V) = 0.075$. These positions and luminositites are sent to the {\sc SkyMaker} software package \citep{bertin:09}, which generates the
thumbnail images\footnote{By default {\sc SimClust} provides $UBVRIJHK$ magnitudes to {\sc SkyMaker}, which then produces images in these passbands. 
We made a small modification to the software in order to produce $g$-band magnitudes (and images), which we
calculated according to the inverse of the transformation equations in Section \ref{ss:phottrans}.}. {\sc Skymaker} also
requires a model PSF, which we generated using the {\sc iraf} {\it psf} and {\it seepsf} tasks assuming a Gaussian profile of FWHM $\sim 0.7\arcsec$,
corresponding to the mean $g$-band stellar profile in PAndAS. We further specified the remainder of the {\sc SkyMaker} parameters
to have values appropriate for CFHT/MegaCam and PAndAS. Finally, we also employed the ability of {\sc SkyMaker} to randomly
add field stars across each thumbnail; we tweaked the relevant parameters to match, empirically, the range of field densities observed
locally about our PAndAS GCs. Figure \ref{f:fakeclusters} shows $1\arcmin \times 1\arcmin$ central cut-outs from a handful of representative 
artificial GC images, across the luminosity-concentration plane.

\begin{figure}
\begin{center}
\includegraphics[width=80mm]{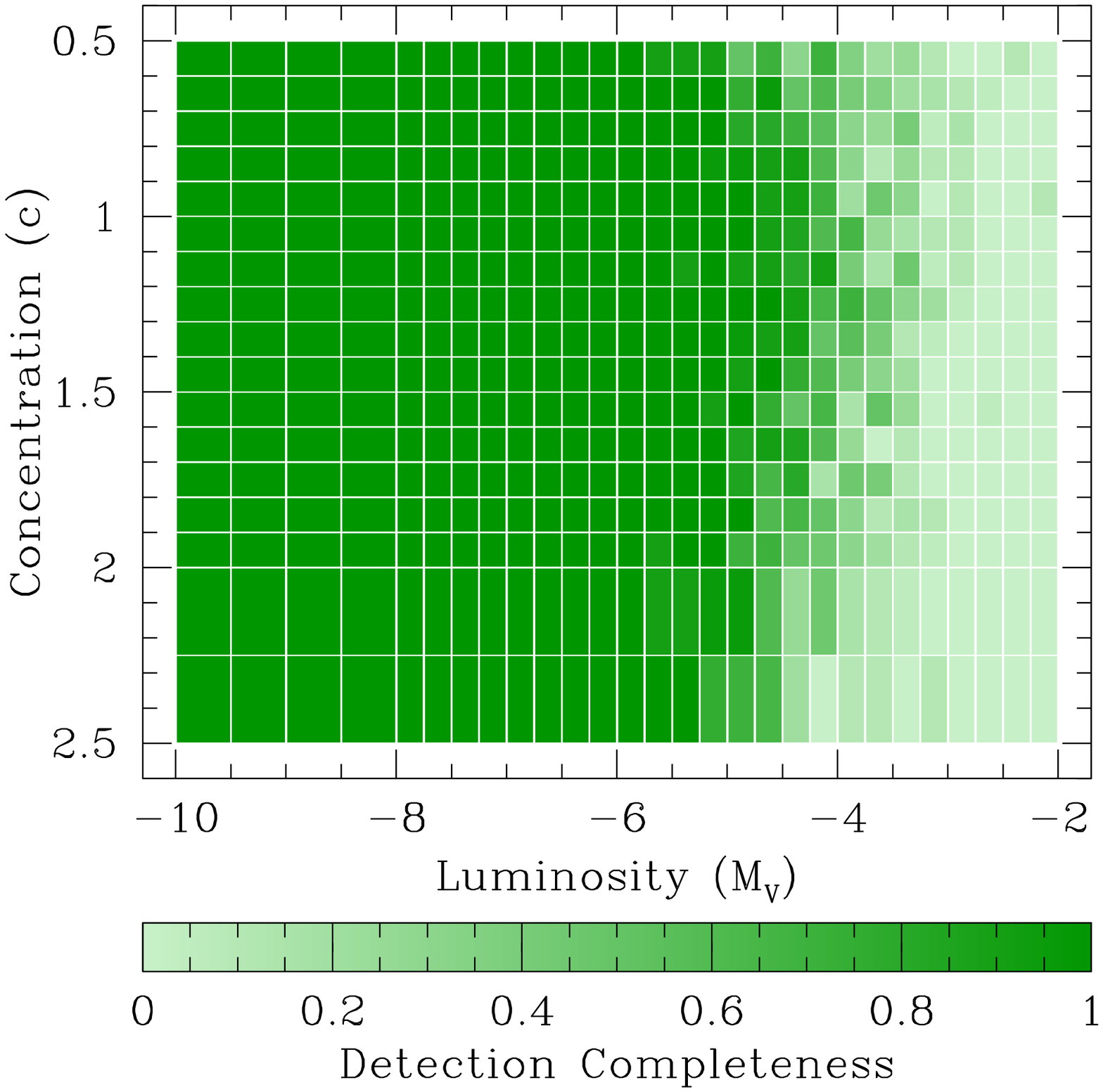}
\caption{Detection completeness as a function of luminosity and concentration from our artificial cluster tests.}
\label{f:completeness1}
\end{center}
\end{figure}
  
\begin{figure}
\begin{center}
\includegraphics[width=80mm]{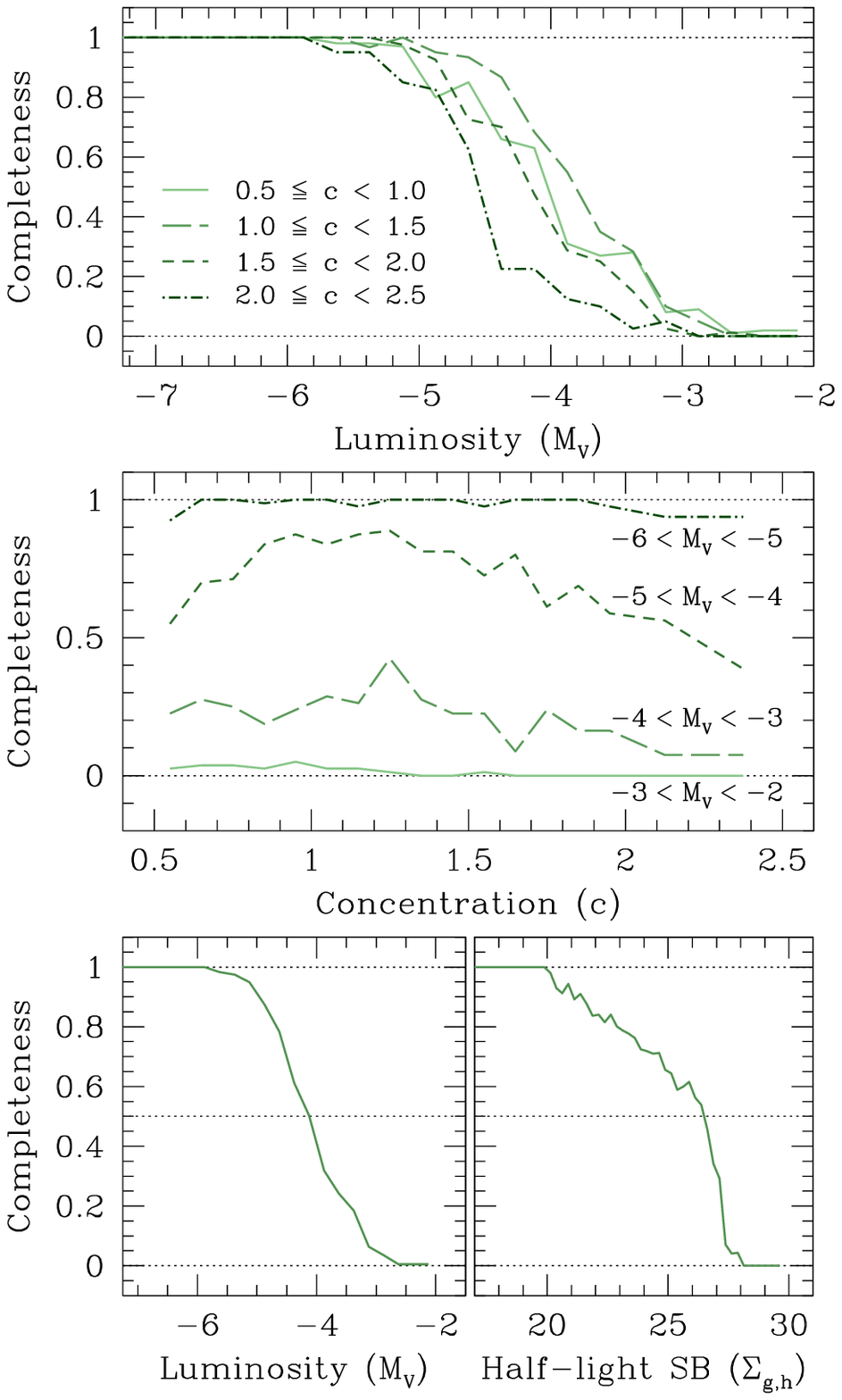}
\caption{{\bf Upper panel:} Detection completeness as a function of cluster luminosity, collapsed into four concentration bins as marked. {\bf Middle panel:} Detection completeness as a function of cluster concentration, collapsed into four luminosity bins as marked. {\bf Lower panels:} Detection completeness across the entire sample of artificial clusters, as a function of luminosity (left) and $g$-band half-light surface brightness $\Sigma_{g,h}$ in magnitudes per square arcsecond (right).}
\label{f:completeness2}
\end{center}
\end{figure}

Once all the artificial GC images had been generated, the order was randomised and the full set supplied to APH for inspection.
To ensure a completely blind test, no accompanying information on individual cluster properties was provided. 
Once the inspection was complete, the classifications
(a simple yes or no for each object) were returned to ADM for analysis. Figures \ref{f:completeness1} and \ref{f:completeness2}
show the results. Our survey is complete to at least $M_{V} = -6$ irrespective of cluster structure; fainter than this, there is a weak
but noticeable dependence on concentration. Peak detectability occurs for clusters with $c \sim 1.25$; there is a gradual
fall-off for concentrations within $\pm 0.75$ of this value, and a greater fall-off for very concentrated clusters with $c \geq 2$.
Whereas our catalogue is $> 95\%$ complete down to $M_{V} \sim -5$ for GCs with $c < 2$, it is only $ \sim 80\%$ 
complete at $M_{V} = -5$ for objects with $c \geq 2$. The main reason for this is that, except for the most diffuse examples, 
GCs in PAndAS are predominantly recognisable as a group of resolved giant stars surrounding an unresolved, or partially resolved, core.
For the most concentrated systems the resolved halo vanishes with decreasing luminosity, leaving just a small central core that is
indistinguishable from a foreground star or compact background galaxy. This effect is clearly visible in Figure \ref{f:fakeclusters}.

Considering the sample as a whole, our $50\%$ completeness level occurs at $M_{V} = -4.1$. The effect of differing cluster 
structures is to move this level by a few tenths of a magnitude in either direction about the mean value.
The $50\%$ completeness levels for GCs with $0.5 \leq c < 1.0$ and $1.5 \leq c < 2.0$ match the mean level very closely.
For those objects with $1.0 \leq c < 1.5$ the $50\%$ level moves somewhat fainter to $M_{V} = -3.8$, while for the most
compact clusters with $2.0 \leq c < 2.5$ the $50\%$ level is substantially brighter at $M_{V} = -4.6$. Note that irrespective
of structure, there is essentially no chance of detecting GCs with $M_{V} \ga -3.2$ in PAndAS imaging. These limits are 
reflected in our real data, where our faintest GC has $M_{V} = -4.06$ (PA-45). Between $-4 < M_{V} < -3$, we expect to 
detect about 20\% of any clusters that are present (see the middle panel of Figure \ref{f:completeness2}). However, we 
found none -- suggesting that (i) few such GCs exist in the halo of M31, and/or (ii) the PAndAS imaging data are somewhat 
more demanding than the synthetic GC data used to estimate the completeness, and/or (iii) the mild selection bias we 
described above for very low surface brightness clusters has pushed our faint-end completeness limits too low by a few tenths of a magnitude.

It is informative to consider our completeness limits in terms of the half-light surface brightness $\Sigma_{g,h}$ 
(that is, the mean $g$-band surface brightness within the cluster half-light radius $r_h$). Here, the fall-off is very sharp 
-- our mean $50\%$ limit occurs at $\Sigma_{g,h} = 26.5$ mag arcsec$^{-2}$, and there is essentially no chance of detecting 
clusters with $\Sigma_{g,h} > 27.2$ mag arcsec$^{-2}$.  Again, this ties in well with our detections. We have just one GC with 
$\Sigma_{g,h} \approx 27$ mag arcsec$^{-2}$ (PA-03), and only another four with $\Sigma_{g,h} \ga 26$ mag arcsec$^{-2}$.  
As usual, we are assuming $\mu = 24.47$ for M31, and a typical foreground extinction of $E(B-V) = 0.075$. Under such assumptions, 
our surface brightness limits recast in terms of $V$ (i.e., $\Sigma_{V,h}$) would be $ \sim 0.35$ mag arcsec$^{-2}$ brighter. 
Note that it is unnecessary to consider the effect of cluster structure on these limits because the faint end of the function 
($\Sigma_{g,h} > 26$ mag arcsec$^{-2}$) samples only diffuse ($r_{h} > 10$\ pc) and relatively low luminosity ($M_{V} > -6$) GCs.

\section{Results \& Analysis}
\label{analysis}
In this section we explore the properties of the enlarged M31 halo GC system, using the new clusters described above and exploiting our analysis of completeness which was not available in any of our previous work (e.g., Hux11).
As in Hux11 we study the ensemble photometric properties of the M31 GC system and compare them to those of the GC system of the Milky Way (MW). When taking photometry and structural measurements from the present paper, we only include those clusters which have a quality flag of either `A' or `B' (see Table \ref{t:phot}). We also exclude the two candidate GCs from our analysis.

In the analysis that follows, we supplement our catalogue of outer halo GCs with confirmed GCs from the most recent revision of the RBC (almost all of which are at $R_{\rm proj} < 25$ kpc). Since Hux11 there have been a number of significant changes to the RBC --  Hux11 used version 3.5, and this has now been updated to version 5. In particular, the latest version adds the photometry of \citet{Fanetal10} and \citet{Peacocketal10}, and the spectroscopy of \citet{Caldwelletal09}. 

The sample of M31 GCs we take from the RBC is defined in a manner comparable to that used in  Hux11, and exploits a number of flags provided in the RBC that help classify the characteristics of the GCs. We only use those objects for which the RBC flag `f' is set to either 1 or 8 (indicating confirmed compact and extended GCs respectively -- the extended clusters all appear to be old metal-poor systems, so we treat them equally). We thus effectively exclude all objects in the RBC V5 that do not have imaging or spectroscopy confirming their status as GCs.  M31 possesses a population of younger clusters, predominantly set against the galactic disk, which we also exclude as there are no comparable clusters in our MW sample. This was achieved by ensuring the flag `yy' is 0 -- indicating clusters that are ``not young" according the data of \citet{FusiPeccietal05}, based on the $(B-V)$ colour or the strength of the H$\beta$ spectral index. Additional young clusters are excluded by removing objects for which the flag `ac' is 1 or 2 (which indicate an age estimate of less than 1 Gyr, or $\sim 1-2$ Gyr, respectively, drawing on the spectroscopy of \citealt{Caldwelletal09}); and for which the flag `pe' is not 0, 1 or 2 (based on \citealt{Peacocketal10}, who use broadband colours to identify likely young clusters). 

These selection criteria leave a sample of 425 GCs from the RBC V5, which is actually fewer than the RBC V3.5 sample employed in Hux11 even though the catalogue now includes the 40 new GCs we presented in that paper. This reduction is due to the large number of objects that are either now known not to be GCs, or are now classified as being ``young".

Many of the GCs in version 5 of the RBC also have new $E(B-V)$ values compared to version 3.5. In addition to those of \citet{Fanetal10}, which were available for  Hux11, new values have been derived by \citet{Caldwelletal11} from their spectroscopy. If any particular GC is in both \citet{Fanetal10} and \citet{Caldwelletal11}, we take the mean of the two $E(B-V)$ values given for that cluster. If a GC within 25 kpc of the centre of M31 has no colour excess from either of these sources, we apply an average value of $E(B-V)$ derived from the medians of both samples. However, for those GCs with $R_{\rm proj} > 25$ kpc, we use the $E(B-V)$ values derived from the \citet{Schlegeletal98} dust maps, as additional internal M31 reddening towards these objects is probably negligible. Note that we employ  the updated reddening values from \citet{Schlaflyetal11} as listed in Table \ref{t:phot}.

Data on the MW GC system have also been updated since  Hux11 was written. We now use the current version (dated December 2010) of the McMaster catalogue \citep[see][]{Harrisetal96}\footnote{http://www.physics.mcmaster.ca/Globular.html}, although there are no major changes since the previous catalogue. In the analysis that follows we focus on the photometric properties of the GCs, derived from their observed magnitudes and colours.  With this in mind, we exclude from our plots the 28 (of 157 total) MW GCs listed in the McMaster catalogue as having $E(B-V) > 1.0$, as $A_{\lambda}$ is not constant at high extinction. These objects are, in any case, almost all rather poorly studied.

Analysis of the spatial layout of the M31 GCs, and their relationship to stellar substructures in the halo, will be addressed in detail in an accompanying paper in this series (Mackey et al. 2014, in prep).

\subsection{GC luminosity function}

\begin{figure}
 \centering
 \includegraphics[angle=0,width=90mm]{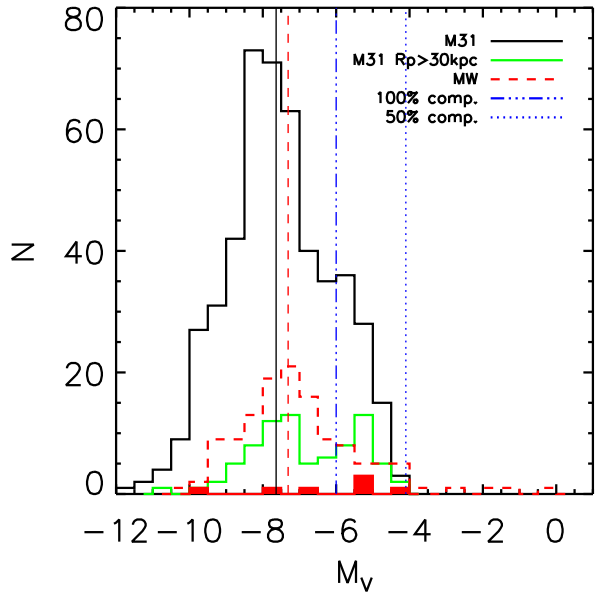}
 \vspace{2pt}
 \caption{Histogram of $M_{V}$, showing the distribution for all M31 GCs, taking the additions and updates from this paper into account (black solid line), and the M31 GCs with a projected galactocentric radius $R_{\rm proj} > 30$ kpc (green solid line) compared to the MW (red  line). The solid black and dashed red vertical lines indicate the median values for the M31 and MW GC systems respectively. The solid red regions show the GCs associated with the Sagittarius dwarf galaxy. The completeness limits for our PAndAS GC search are also shown (blue vertical lines).}\label{Fi:histo_MV0}
\end{figure}

\begin{figure}
 \centering
 \includegraphics[angle=0,width=85mm]{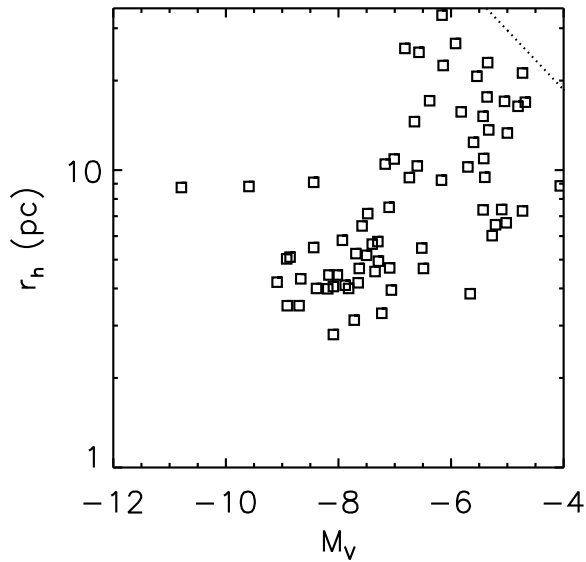}
 \vspace{2pt}
 \caption{Plot of $M_{V}$ against  log$_{10}$($r_{h}$) for M31 GCs in the outer halo, with a projected distance greater than 30 kpc. The more luminous clusters ($M_{V} < -7.5$) are relatively compact, whilst the fainter clusters span a broad range of effective radii. Recall that size measurements for any GC with $r_h \la 8-10$ pc are upper limits. The dashed line shows the location of our 50\% completeness limit.} \label{Fi:plot_MV0_Rh}
\end{figure}

The M31 globular cluster luminosity function (GCLF) is shown in Figure \ref{Fi:histo_MV0}. The median value of  $M_{V}$ for M31 is $-7.6$,  compared to the $-7.9$ that we found in  Hux11; that for the MW sits at $-7.3$, the same as in Hux11 despite the various minor updates to the McMaster catalogue.

In Hux11 we suggested that the then available data indicated a secondary peak in the M31 GCLF at $M_{V} \sim -6$ mags, although we noted at the time that many of the objects identified by \citet{Kimetal07} as clusters were near this magnitude, but had questionable classifications.  We further found that the secondary peak was visible when both inner halo ($R_{\rm proj} < 25$ kpc) and outer halo ($R_{\rm proj} > 25$ kpc) GCs were considered separately, suggesting that this might indeed be a real feature. 

The new data reveal a more complex situation. With the updated RBC, the second peak for the full M31 sample (black histogram) no longer appears. This is primarily due to the reclassification of many of the \citet{Kimetal07} GCs as stellar contaminants \citep{Peacocketal10}, which reduces the number of confirmed GCs in this magnitude range. However, if we consider only the outer halo clusters (in this case $R_{\rm proj} > 30$ kpc), shown in green in Figure \ref{Fi:histo_MV0}, a bimodality in the GCLF is very clear with peaks at $\sim -7.5$ and $\sim -5.5$.  The fainter secondary peak sits between our 100\% and 50\% completeness limits. Hence it is possible that additional GCs exist around this luminosity, that we have not detected. These would further increase the prominence of the feature, and the location of the peak may shift slightly (probably towards slightly fainter magnitudes).

It is natural to ask about the typical nature of the GCs residing in the secondary peak. \citet{Mackeyetal10a} argue that a substantial fraction (perhaps up to $\sim80$\%) of the M31 GCs with $R_{\rm proj} > 30$ kpc have been accreted into the M31 halo along with their parent dwarf galaxies. Hence, the fainter peak in the GCLF, which is prominent only for the outer halo system, might well be primarily driven by the presence of this type of GC. 

This scenario finds additional support if we consider the MW GCs that are believed to be associated with the Sagittarius dwarf galaxy \citep{LawMajewski10}. Although there are only perhaps eight such GCs (Arp 2, NGC 6715, NGC 5634, Terzan 7, Terzan 8, NGC 5053, Pal 12 and Whiting 1), five of these have luminosities fainter than $M_{V} = -6$.  More generally, \citet{MackeyvandenBergh05} found a similar fainter peak in GCLF of the``young halo" GCs of the MW, which they argued (from indirect evidence) are most likely accreted objects. 

The M31 halo GCs near the secondary peak in the GCLF differ in other ways. A plot of $M_{V}$ against $r_{h}$ for the M31 GCs beyond 30 kpc (Figure \ref{Fi:plot_MV0_Rh}) shows that clusters with a luminosity near the fainter peak of the GCLF span a very broad range of half-light radii, while clusters near the more luminous peak are primarily compact. 

\begin{figure}
 \centering
 \includegraphics[angle=0,width=85mm]{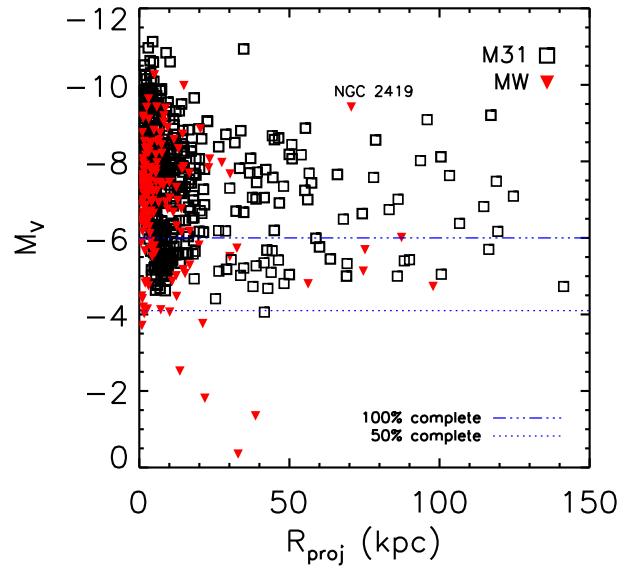}
 \vspace{2pt}
 \caption{Plot of $M_{V}$ against projected galactocentric radius $R_{\rm proj}$, with the completeness limits of our PAndAS search again shown (blue).  In the case of the MW GCs the actual distance  (R$_{\rm gc}$) in this, and subsequent plots, is converted to an ``average projected distance" via the relationship $R_{\rm proj} = R_{\rm gc}\times (\pi/4)$. There are many luminous GCs in M31 at large radii, but a similarly abundant population is not seen in the MW. }\label{Fi:plot_R_MV0}
\end{figure}

Our GCLF for the outer M31 halo suggests that there is a substantial population of luminous GCs outside $R_{\rm proj} = 30$ kpc.  A plot of absolute magnitude against projected radius (Fig. \ref{Fi:plot_R_MV0}) makes this more explicit -- our new PAndAS GC search has yielded many more luminous clusters in the outer halo compared to Hux11. Note that the MW GCs are plotted with an  ``average projected distance", via the relationship $R_{\rm proj} = R_{\rm gc}\times (\pi/4)$, to make the published Galactocentric distances more directly comparable with the projected values of the M31 GCs.  The number of luminous GCs at large galactocentric radii is in striking contrast to the situation seen in the MW. The only GC in the MW which is comparably luminous and also at a large distance from the Galactic centre is the unusual object NGC 2419. 

There is a population of very low luminosity clusters ($M_{V} > -4$) in the MW, which we would likely not see in the PAndAS data -- if they were present -- as at the M31 distance they lie well below our 50\% completeness limit. However, in the MW, these faint GCs are found at moderately large (projected) galactocentric radii, suggesting that deeper imaging in the future may indeed reveal such objects in the halo of M31.

\subsection{GC colour distribution}

\begin{figure}
 \centering
 \includegraphics[angle=0,width=90mm]{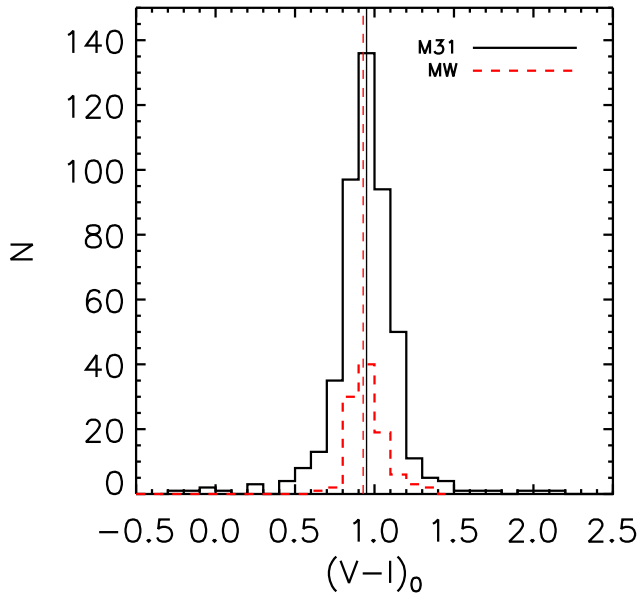}
 \vspace{2pt}
 \caption{Histogram of $(V-I)_{0}$. The vertical lines show the median values for the full sample (solid) and the MW (red).} \label{Fi:histo_VI0}
\end{figure}

The distribution of $(V-I)_{0}$ colours (Figure \ref{Fi:histo_VI0}) shows almost no difference to that found by  Hux11. The median $(V-I)_{0}$ values for GCs in M31 and the MW are almost indistinguishable at 0.95 and 0.93, respectively.

\begin{figure}
 \centering
 \includegraphics[angle=0,width=85mm]{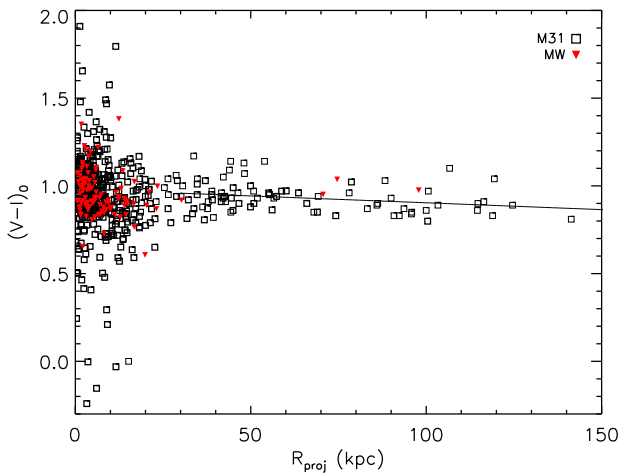}
 \vspace{2pt}
 \caption{Plot of $(V-I)_0$  against $R_{\rm proj}$. The black line shows a linear fit to the M31 GCs with $R_{\rm proj} > 30$ kpc. }\label{Fi:plot_VI0}
\end{figure}

When viewed as a function of galactocentric radius (Figure \ref{Fi:plot_VI0}), the results are again similar to those reported by  Hux11. In that paper, we found a flat colour-radius relation for GCs in the outer halo. The new data are consistent with this, exhibiting only a marginal slope of $-0.0007 \pm 0.0004$ magnitudes per kpc for the GCs beyond 30 kpc. This uncertainty is reflected in spectroscopic results for a limited subset of M31 halo GCs, under the assumption that changes in integrated colour are largely a result of changing metallicity. \citet{Fanetal11} found a small gradient (to decreasing metallicity at larger galactocentric distances) for GCs with $R_{\rm proj} > 25$ kpc, while \citet{Coluccietal12} find a nearly constant metallicity for GCs with $R_{\rm proj} > 20$ kpc. Both these studies used previously published GCs, and although their sample extend to over 100 kpc, they have very few clusters in the distant halo. Our data greatly increase the sample size, improving the robustness of the result.

It is noticeable that although the $(V-I)_{0}$ colours of the full M31 and MW GC systems are almost identical, for the outer halo the MW GCs typically appear slightly redder than the bulk of the M31 GCs at a comparable distance -- although as there are very few MW GCs at these large galactocentric radii, it is difficult to draw firm conclusions. 

\subsection{Cluster sizes}
One result that differs significantly from that seen by  Hux11 concerns the distribution of half-light radii for outer halo M31 GCs. In  Hux11 we  suggested this may take a bimodal form, with one peak corresponding to the typical sizes of traditional compact GCs ($r_h \sim 3-5$ pc) and the other at much larger $r_h \ga 15$ pc. \citet{WangMa13} also reported a size bimodality for M31 GCs at $R_{\rm proj} > 40$ kpc, but they used a small sample of clusters from Hux08 which were also included in our Hux11 analysis. We can now address this question definitively with our larger halo GC sample. 

For comparison purposes we also assemble a set of (inner) M31 GCs with recent size measurements in the RBC. The largest RBC sample comes from the compilation of \citet{Peacocketal09,Peacocketal10}, which we supplement with measurements from \citet{Barmbyetal07} when only the latter is available. Note that \citet{Peacocketal09} provided a careful demonstration that their GC size measurements showed excellent consistency with those derived from {\it HST} imaging by \citet{Barmbyetal07}.

We remind the reader of the reliability of our determination of cluster effective radius as described in Section \ref{photometry}. For clusters smaller than $r_{h} \sim 8 - 10$ pc, the size measurements are significantly affected by the seeing profile (see Figure \ref{f:rhfake}) and we thus over-estimate their values, typically by $\approx 20-30\%$. However, the relative ordering of such GCs by size ought to be largely correct. As described previously, some of the GCs we report measurements for in Table \ref{t:phot} also have sizes measured from {\it HST} imaging by \citet{Tanviretal12}. We take these in preference where available, to try and minimise the effects of this issue.

\begin{figure}
 \centering
 \includegraphics[angle=0,width=85mm]{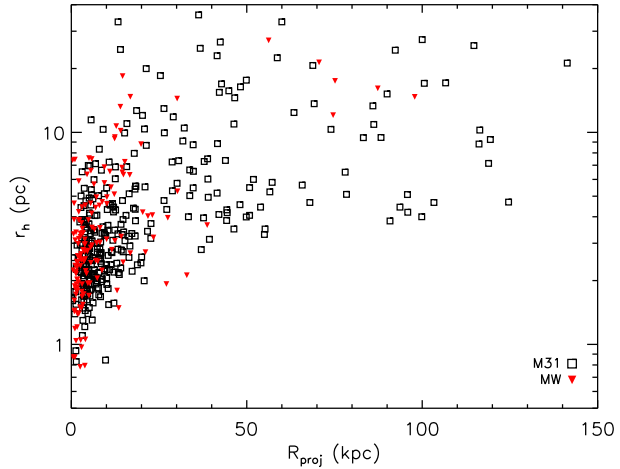}
 \vspace{2pt}
 \caption{Plot of log$_{10}$($r_{h}$)  against $R_{\rm proj}$. There is an apparently continuous range of half-light sizes at large galactocentric radii.}\label{Fi:m31_mw_r_rh_p3}
\end{figure}

 In Figure \ref{Fi:m31_mw_r_rh_p3} we show $r_{h}$ for M31 and MW GCs against projected distance from the centre of their host galaxy. The apparent bimodality in cluster size at large galactocentric radii for M31 GCs, as noted in  Hux11, has now vanished. There are many clusters with $r_{h}$ between 5 and 20 pc, in which range the size distribution appears rather evenly spread at all galactocentric radii outside $\sim 10$ kpc. The original observation of bimodality may have been partly due to the difficulty in measuring accurate GC sizes from our INT/WFC data. We observe significant changes in the inferred sizes of some objects moving to the superior PAndAS data -- for example, the clusters H7, H8 and H15 went from our default value for ``compact" GCs of 4.5 pc, used in Hux11, to $\sim$10 pc, while by contrast, some of the more extended clusters have smaller sizes measured from the PAndAS data than those obtained by Hux11. 

The apparently even spread of cluster sizes larger than $\sim 5-10$ pc in the M31 halo strongly suggests that the extended clusters first identified by \citet{Huxoretal05} are simply objects selected from the upper tail of the GC size distribution. This is consistent with their constituent stellar populations, which appear indistinguishable from those observed in typical metal-poor compact GCs \citep[see e.g.,][]{Mackeyetal06,Mackeyetal07}. It is also noticeable from Figure \ref{Fi:m31_mw_r_rh_p3} that the largest clusters observed in the remote MW halo are comparable to the sizes of many of the more extended clusters in M31 -- that is, there do appear to be a few counterparts of the M31 extended clusters seen in the MW halo. The largest M31 clusters have greater $r_h$ than any GCs found in the MW halo, but this is perhaps not surprising given the much more numerous M31 halo GC population.

\begin{figure}
 \centering
 \includegraphics[angle=0,width=85mm]{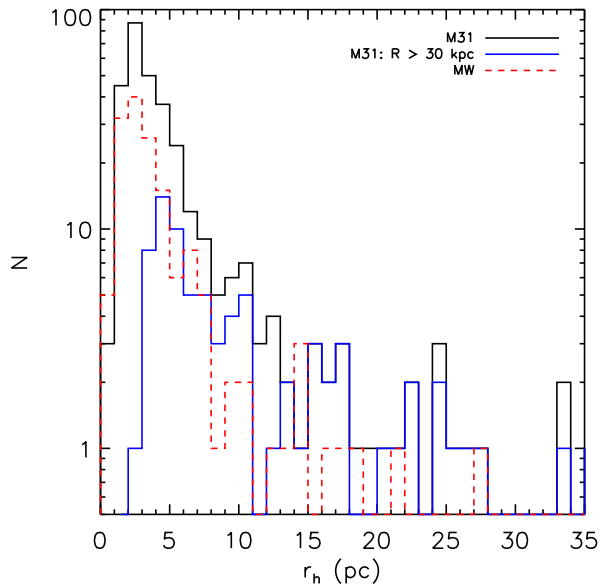}
 \vspace{2pt}
 \caption{Histogram (logarithmic in $N$) of  $r_{h}$. For large galactocentric radii (blue line), the distribution of half-light radii is considerably flatter than for the full M31 sample.}\label{Fi:m31_mw_rh_log_hist_p3}
\end{figure}

Figure \ref{Fi:m31_mw_rh_log_hist_p3} shows a histogram of $r_h$ for M31 and MW GCs. Those for the full systems appear to share a very similar shape. However it is notable that the distribution of $r_h$ for M31 GCs with a galactocentric distance $>$ 30 kpc is quite unlike that for the full M31 sample. Even taking into account the tendency for our PAndAS measurements to over-estimate the sizes of GCs with $r_h \la 8-10$ pc, the distribution of half-light radii for clusters more than 30 kpc from M31 would still be considerably flatter than that of the full sample. That is, the ratio of the number of GCs with $r_h$ above $8-10$ pc to those with $r_h$ below this level is substantially greater for M31 GCs with $ R_{\rm proj} > 30$ kpc than for the full M31 sample. A similar pattern is seen in the MW, albeit at lower significance due to the smaller numbers of clusters involved \citep[e.g.,][]{MackeyGilmore04,MackeyvandenBergh05}. It is unclear to what extent this situation reflects the lower tidal fields in the outer regions of M31 (and indeed the MW), or whether it is due to the likely origin of many of these GCs in accreted dwarf galaxies \citep[see also the discussion in][]{DaCostaetal09,Hwangetal11}.

\section{Conclusions}

In this paper we present the final catalogue of M31 halo GCs from the PAndAS survey. Of these, 57 were identified by our usual method of visually searching the new image data, and one further cluster was found by a code searching for faint dwarf galaxies. Our catalogue represents the first detailed and uniform census of GCs across nearly the full extent of the M31 halo.  We find numerous clusters with very large projected galactocentric radii ($R_{\rm proj} \ga 100$ kpc), reflecting the huge spatial extent of the M31 GC system.

We located a few additional GCs by revisiting outer halo candidates listed in the RBC. We found that three such candidates are indeed GCs, while one is a H{\sc ii} region with a possible embedded young cluster; and we also located one further new discovery that serendipitously falls near a star that was the source of the RBC entry. In addition, we found that three ``definite" outer halo GCs listed in the RBC are not clusters after all. Finally, we confirm that ten of the 17 ``high-confidence" SDSS clusters listed by \citet{dtz:13} are indeed GCs, based on our higher-quality PAndAS imaging. However, only one of their 42 ``candidate" objects that we were able to examine was found to be a cluster. 

Experiments with artificial clusters suggest that our GC survey is complete down to a cluster luminosity of $M_{V} = -6.0$, and has 50\% completeness limit at roughly $M_{V} \approx -4.1$ . Our analysis indicates that an additional $\sim 3 - 5$ clusters may lie undiscovered within the area covered by PAndAS imaging (i.e., within $\approx 150$ kpc of M31), due to small gaps in the survey coverage. We cannot rule out that there may also be many very faint clusters with $M_{V} \ga -4$ that we are unable to detect using PAndAS. 

We used the PAndAS imaging to measure luminosities, colours and sizes for all known M31 GCs outside $ R_{\rm proj} = 25$ kpc. The results of this process confirm most of the findings from Hux11 with a much larger sample. The bimodality of the luminosity function constructed using M31 halo GCs with $R_{\rm proj} > 30$ kpc is perhaps the most notable feature. This bimodality is not seen in the LF constructed using more central clusters, and we suggest it may be a consequence of the dwarf galaxy accretion history of the outer M31 halo. The colours of the halo GCs show only a marginally significant shallow gradient with projected radius, while the distribution of half-light radii for the M31 halo GCs reveals an apparently continuous spread of cluster sizes, rather than the bimodality suggested by previous studies that used much smaller samples and shallower imaging.

Many of the new GCs described here have already been followed up by the PAndAS collaboration. For example, a large fraction of these objects is included in the studies of \citet{Veljanoskietal13a} and the companion paper to the present work by \citet{Veljanoskietal14}, where radial velocities have been used to explore the kinematics of the M31 outer halo GC system.

Individual clusters have also proved of interest. In \citet{Mackeyetal13b} we investigated  two of the new PAndAS GCs (PA-7  and PA-8), which are almost certainly associated with a prominent halo substructure known as the South-West Cloud \citep[see][]{Lewisetal13,Bateetal14}. These objects appear to be at least 2 Gyr younger than the oldest MW GCs, and thus fit with the trends identified by \citet{Perinaetal12}, and show strong similarities to the supposedly-accreted ``young halo" clusters in the MW \citep{MackeyvandenBergh05}. 

Our new clusters also provide a substantial number of GCs which exhibit properties unlike those studied in the MW. Examples include the few very most extended clusters, and the luminous, compact clusters found in the far halo of M31. Some of the new GCs may be of major interest. For example, PA-48 has a structure and ellipticity that may be more akin to a very faint dwarf galaxy than a typical globular cluster \citep[see][]{Mackeyetal13a}.

{\it HST} imaging reaches to below the horizontal branch at the distance of M31 in a just a couple of orbits -- although it is a challenge to go much deeper. \citet{Brownetal04} required a total of  3.5 days of exposure time to reach to 1.5 mag below the old main sequence turn-off of the M31 globular cluster SKHB 312. However, this situation will change with the launch of {\em JWST}, which should be able to reach the main sequence turn-off for M31 GCs with manageable exposure times, allowing us to investigate the GC system of M31 in a manner comparable to our current understanding of the Galactic GC system. With low contaminating backgrounds, the GCs presented here will be ideal targets for such studies.

\section*{Acknowledgments}
We would like to thank the referee, Flavio Fusi Pecci, for a detailed and constructive report that helped improve this paper. We also appreciate the careful reading by Luciana Federici and Silvia Galleti.

ADM is grateful for support by an Australian Research Fellowship (Grant DP1093431) from the Australian Research Council.  AMNF, APH, and ADM acknowledge support by a Marie Curie Excellence Grant from the European Commission under contract MCEXT-CT-2005-025869, during which this work was initiated. 

This work is based on observations obtained with MegaPrime/MegaCam, a joint project of CFHT and CEA/DAPNIA, at the Canada-France-Hawaii Telescope (CFHT) which is operated by the National Research Council (NRC) of Canada, the Institute National des Sciences de l'Univers of the Centre National de la Recherche Scientifique of France, and the University of Hawaii.

\label{lastpage}

\begin{thebibliography}{99}

\bibitem[\protect\citeauthoryear{Alves-Brito et 
al.}{2009}]{AlvesBritoetal09} Alves-Brito A., Forbes D.~A., Mendel 
J.~T., Hau G.~K.~T., Murphy M.~T., 2009, MNRAS, 395, L34 

\bibitem[\protect\citeauthoryear{Barmby et al.}{2000}]{Barmbyetal00}
  Barmby P., Huchra J.~P., Brodie J.~P., Forbes D.~A., Schroder L.~L.,
  Grillmair C.~J., 2000, AJ, 119, 727

\bibitem[\protect\citeauthoryear{Barmby \& Huchra}{2001}]{barmby:01}
  Barmby P., Huchra J.~P., 2001, AJ, 122, 2458
 
\bibitem[\protect\citeauthoryear{Barmby et al.}{2007}]{Barmbyetal07} 
Barmby P., McLaughlin D.~E., Harris W.~E., Harris G.~L.~H., Forbes D.~A., 
2007, AJ, 133, 2764

\bibitem[\protect\citeauthoryear{Bate et al.}{2014}]{Bateetal14}
  Bate N.F., et al., 2014, MNRAS, 437, 3362

\bibitem[\protect\citeauthoryear{Battistini et al.}{1987}]{Battistinietal87}
  Battistini P., Bonoli F., Braccesi A., Federici L., Fusi Pecci F., Marano B., Borngen, F., 1987, A\&AS, 64, 447
  
\bibitem[\protect\citeauthoryear{Bertin}{2009}]{bertin:09}
  Bertin E., 2009, MmSAI, 80, 422
 
 \bibitem[\protect\citeauthoryear{Brown et al.}{2004}]{Brownetal04} 
Brown T.~M., Ferguson H.~C., Smith E., Kimble R.~A., Sweigart A.~V., 
Renzini A., Rich R.~M., VandenBerg D.~A., 2004, ApJ, 613, L125

\bibitem[\protect\citeauthoryear{Caldwell et 
al.}{2009}]{Caldwelletal09} Caldwell N., Harding P., Morrison H., Rose 
J.~A., Schiavon R., Kriessler J., 2009, AJ, 137, 94 

\bibitem[\protect\citeauthoryear{Caldwell et 
al.}{2011}]{Caldwelletal11} Caldwell N., Schiavon R., Morrison H., 
Rose J.~A., Harding P., 2011, AJ, 141, 61 

\bibitem[\protect\citeauthoryear{Chies-Santos et 
al.}{2011}]{ChiesSantosetal11} Chies-Santos A.~L., Larsen S.~S., Kuntschner H., Anders P., Wehner E.~M., Strader J., Brodie J.~P., Santos J.~F.~C., 2011, A\&A, 525, A20 

\bibitem[\protect\citeauthoryear{Cockcroft et 
al.}{2011}]{Cockcroftetal11} Cockcroft R., et al., 2011, ApJ, 730, 112 
 
\bibitem[\protect\citeauthoryear{Colucci, Bernstein, 
\& Cohen}{2012}]{Coluccietal12} Colucci J., Bernstein R.~A., Cohen J., 2012, Proceedings of the XII International Symposium on Nuclei in the Cosmos,  

\bibitem[\protect\citeauthoryear{Conn et al.}{2012}]{Connetal12}
Conn A.~R., et al., 2012, ApJ, 758, 11

\bibitem[\protect\citeauthoryear{Crampton et
    al.}{1985}]{Cramptonetal85} Crampton D., Cowley A.~P., Schade D.,
  Chayer P., 1985, ApJ, 288, 494
 
\bibitem[\protect\citeauthoryear{Da Costa et al.}{2009}]{DaCostaetal09}
 Da Costa G. S., Grebel E. K., Jerjen H., Rejkuba M., Sharina M. E., 2009, AJ, 137, 4361
 
\bibitem[\protect\citeauthoryear{Deveikis et al.}{2008}]{deveikis:08}
  Deveikis V., Narbutis D., Stonkut\.{e} R., Brid\v{z}ius A., Vansevi\v{c}ius V., 2008, BaltA, 17, 351

\bibitem[\protect\citeauthoryear{di Tullio Zinn \& Zinn}{2013}]{dtz:13}
  di Tullio Zinn G., Zinn R., 2013, AJ, 145, 50
  
\bibitem[\protect\citeauthoryear{Dotter et al.}{2010}]{dotter:10} 
  Dotter A., et al., 2010, ApJ, 708, 698

\bibitem[\protect\citeauthoryear{Elson \& Walterbos}{1988}]{Elson88}
  Elson R.~A., Walterbos R.~A.~M., 1988, ApJ, 333, 594

\bibitem[\protect\citeauthoryear{Fan et al.}{2008}]{Fanetal08} 
Fan Z., Ma J., de Grijs R., Zhou X., 2008, MNRAS, 385, 1973 

\bibitem[\protect\citeauthoryear{Fan, de Grijs, 
\& Zhou}{2010}]{Fanetal10} Fan Z., de Grijs R., Zhou X., 2010, ApJ, 725, 200 

\bibitem[\protect\citeauthoryear{Fan et al.}{2011}]{Fanetal11} 
Fan Z., Huang Y.-F., Li J.-Z., Zhou X., Ma J., Wu H., Zhang T.-M., Zhao 
Y.-H., 2011, RAA, 11, 1298 

\bibitem[\protect\citeauthoryear{Federici et 
al.}{2012}]{Federicietal12} Federici L., Cacciari C., Bellazzini M., Fusi Pecci F., Galleti S., Perina S., 2012, A\&A, 544, A155 

\bibitem[\protect\citeauthoryear{Forbes et al.}{2011}]{forbesetal11} 
Forbes D.~A., Spitler L.~R., Strader J., Romanowsky A.~J., Brodie J.~P., 
Foster C., 2011, MNRAS, 413, 2943 

\bibitem[\protect\citeauthoryear{Forte, Vega, 
\& Faifer}{2012}]{forteetal12} Forte J.~C., Vega E.~I., Faifer F., 2012, MNRAS, 421, 635 

\bibitem[\protect\citeauthoryear{Fusi Pecci et
    al.}{2005}]{FusiPeccietal05} Fusi Pecci F., Bellazzini M., Buzzoni
  A., De Simone E., Federici L., Galleti S., 2005, AJ, 130, 554

\bibitem[\protect\citeauthoryear{Galleti et al.}{2004}]{galleti:04} 
  Galleti S., Federici L., Bellazzini M., Fusi Pecci F., Macrina S., 2004, A\&A, 416, 917

\bibitem[\protect\citeauthoryear{Galleti et al.}{2009}]{galleti:09} 
Galleti S., Bellazzini M., Buzzoni A., Federici L., Fusi Pecci F., 2009, A\&A, 508, 1285

\bibitem[\protect\citeauthoryear{Georgiev, Goudfrooij, 
\& Puzia}{2012}]{Georgievetal12} Georgiev I.~Y., Goudfrooij P., Puzia T.~H., 2012, MNRAS, 420, 1317 

\bibitem[\protect\citeauthoryear{Harris}{1996}]{Harrisetal96}
  Harris W.E., 1996, AJ, 112, 1487

\bibitem[\protect\citeauthoryear{Huchra, Brodie, \&
    Kent}{1991}]{Huchraetal91} Huchra J.~P., Brodie J.~P., Kent S.~M.,
  1991, ApJ, 370, 495

\bibitem[\protect\citeauthoryear{Huxor et al.}{2005}]{Huxoretal05}
  Huxor A.P., Tanvir N.R., Irwin M.J., Ibata R., Collett J.L., Ferguson A.M.N., Bridges T., Lewis G.F., 2005, MNRAS, 360, 1007

\bibitem[\protect\citeauthoryear{Huxor et al.}{2008}]{Huxoretal08}
  Huxor A.P., Tanvir N.R., Ferguson A.M.N., Irwin M.J., Ibata R.A., Bridges T., Lewis G.F., 2008, MNRAS, 385, 1989

\bibitem[\protect\citeauthoryear{Huxor et al.}{2009}]{Huxoretal09} 
Huxor A., Ferguson A.~M.~N., Barker M.~K., Tanvir N.~R., Irwin M.~J., 
Chapman S.~C., Ibata R., Lewis G., 2009, ApJ, 698, L77 

\bibitem[\protect\citeauthoryear{Huxor et al.}{2011}]{Huxoretal11}
  Huxor A.P., et al., 2011, MNRAS, 414, 770

\bibitem[\protect\citeauthoryear{Hwang et al.}{2011}]{Hwangetal11}
  Hwang N., Lee M.G., Lee J.C., Park W.-K., Park H.S., Kim S.C., Park J.-H., 2011, ApJ, 738, 58

\bibitem[\protect\citeauthoryear{Ibata et al.}{2007}]{Ibataetal07} 
Ibata R., Martin N.~F., Irwin M., Chapman S., Ferguson A.~M.~N., Lewis 
G.~F., McConnachie A.~W., 2007, ApJ, 671, 1591 

\bibitem[\protect\citeauthoryear{Ibata et al.}{2014}]{Ibataetal13} 
Ibata R.~A., et al., 2014, ApJ, 780, 128

\bibitem[\protect\citeauthoryear{King}{1962}]{king:62}
  King I., 1962, AJ, 67, 471
  
  \bibitem[\protect\citeauthoryear{Kim et al.}{2007}]{Kimetal07} 
Kim S.~C., et al., 2007, AJ, 134, 706 

\bibitem[\protect\citeauthoryear{Kroupa}{2001}]{kroupa:01}
  Kroupa P., 2001, MNRAS, 322, 231

\bibitem[\protect\citeauthoryear{Law 
\& Majewski}{2010}]{LawMajewski10} Law D.~R., Majewski S.~R., 2010, ApJ, 718, 1128 

\bibitem[\protect\citeauthoryear{Lewis et al.}{2013}]{Lewisetal13}
  Lewis G.F., et al., 2013, ApJ, 763, 4
  
\bibitem[\protect\citeauthoryear{Ma}{2012}]{Ma12} Ma J., 
2012, RAA, 12, 115 

\bibitem[\protect\citeauthoryear{Mackey \& Gilmore}{2004}]{MackeyGilmore04}
 Mackey A.D., Gilmore G.F., 2004, MNRAS, 355, 504

\bibitem[\protect\citeauthoryear{Mackey 
\& van den Bergh}{2005}]{MackeyvandenBergh05} Mackey A.~D., van den Bergh S., 2005, MNRAS, 360, 631 

\bibitem[\protect\citeauthoryear{Mackey et al.}{2006}]{Mackeyetal06}
  Mackey A.D., et al., 2006, ApJ, 653, L105

\bibitem[\protect\citeauthoryear{Mackey et al.}{2007}]{Mackeyetal07}
  Mackey A.D., et al., 2007, ApJ, 655, L85

\bibitem[\protect\citeauthoryear{Mackey et al.}{2010a}]{Mackeyetal10a}
  Mackey A.D., et al., 2010a, ApJ, 717, L11

\bibitem[\protect\citeauthoryear{Mackey et al.}{2010b}]{Mackeyetal10b}
  Mackey A.D., et al., 2010b, MNRAS, 401, 533

\bibitem[\protect\citeauthoryear{Mackey et al.}{2013a}]{Mackeyetal13a}
  Mackey A.D., et al., 2013a, ApJ, 770, L17  

\bibitem[\protect\citeauthoryear{Mackey et al.}{2013b}]{Mackeyetal13b} 
  Mackey A.D., et al., 2013b, MNRAS, 429, 281 

\bibitem[\protect\citeauthoryear{Marigo et al.}{2008}]{marigo:08}
  Marigo P., Girardi L., Bressan A., Groenewegen M.A.T., Silva L., Granato G.L., 2008, A\&A, 482, 883

\bibitem[\protect\citeauthoryear{Mar\'{i}n-Franch et al.}{2009}]{marinfranch:09}
  Mar\'{i}n-Franch A., et al., 2009, ApJ, 694, 1498

\bibitem[\protect\citeauthoryear{Martin et al.}{2006}]{Martinetal06} 
  Martin N.F., Ibata R.A., Irwin M.J., Chapman S., Lewis G.F., Ferguson A.M.N., Tanvir N., McConnachie A.W., 2006, MNRAS, 371, 1983
  
 \bibitem[\protect\citeauthoryear{Martin et al.}{2013}]{Martinetal13} 
Martin N.~F., Ibata R.~A., McConnachie A.~W., Mackey A.~D., Ferguson 
A.~M.~N., Irwin M.~J., Lewis G.~F., Fardal M.~A., 2013, ApJ, 776, 80 

\bibitem[\protect\citeauthoryear{McConnachie et al.}{2005}]{McConnachieetal05} 
McConnachie A.~W., Irwin M.~J., Ferguson A.~M.~N., Ibata R.~A., Lewis G.~F., Tanvir N., 2005, MNRAS, 356, 979

\bibitem[\protect\citeauthoryear{McConnachie et al.}{2008}]{McConnachieetal08} 
McConnachie A.~W., et al., 2008, ApJ, 688, 1009 

\bibitem[\protect\citeauthoryear{McConnachie et al.}{2009}]{McConnachieetal09} 
McConnachie A.~W., et al., 2009, Nature, 461, 66

\bibitem[\protect\citeauthoryear{Peacock et al.}{2009}]{Peacocketal09} 
Peacock M.~B., Maccarone T.~J., Waters C.~Z., Kundu A., Zepf S.~E., Knigge C., 
Zurek D.~R., 2009, MNRAS, 392, L55 

\bibitem[\protect\citeauthoryear{Peacock et al.}{2010}]{Peacocketal10} 
Peacock M.~B., Maccarone T.~J., Knigge C., 
Kundu A., Waters C.~Z., Zepf S.~E., Zurek D.~R., 2010, MNRAS, 402, 803 

\bibitem[\protect\citeauthoryear{Perina et al.}{2009}]{Perinaetal09} 
Perina S., Federici L., Bellazzini M., Cacciari C., Fusi Pecci F., Galleti S., 2009, A\&A, 507, 1375

\bibitem[\protect\citeauthoryear{Perina et al.}{2011}]{Perinaetal11} 
Perina S., Galleti S., Fusi Pecci F., Bellazzini M., Federici L., Buzzoni A., 2011, A\&A, 531, 155

\bibitem[\protect\citeauthoryear{Perina et al.}{2012}]{Perinaetal12} 
  Perina S., Bellazzini M., Buzzoni A., Cacciari C., Federici L., Fusi Pecci F., Galleti S., 2012, A\&A, 546, A31 

\bibitem[\protect\citeauthoryear{Perrett et al.}{2002}]{Perrettetal02}
  Perrett K.~M., Bridges T.~J., Hanes D.~A., Irwin M.~J., Brodie
  J.~P., Carter D., Huchra J.~P., Watson F.~G., 2002, AJ, 123, 2490

\bibitem[\protect\citeauthoryear{Schlafly et al.}{2011}]{Schlaflyetal11}
  Schlafly E.F., Finkbeiner D.P., 2011, ApJ, 737, 103
  
\bibitem[\protect\citeauthoryear{Schlegel et al.}{1998}]{Schlegeletal98}
  Schlegel D.J., Finkbeiner D.P., Davis M., 1998, ApJ, 500, 525

\bibitem[\protect\citeauthoryear{Searle 
\& Zinn}{1978}]{SearleZinn78} Searle L., Zinn R., 1978, ApJ, 225, 357

\bibitem[\protect\citeauthoryear{Tanvir et al.}{2012}]{Tanviretal12}
  Tanvir N.R., et al., 2012, MNRAS, 422, 162

\bibitem[\protect\citeauthoryear{Veljanoski et 
al.}{2013a}]{Veljanoskietal13a} Veljanoski J., et al., 2013a, ApJ, 768, L33 

\bibitem[\protect\citeauthoryear{Veljanoski et 
al.}{2013b}]{Veljanoskietal13b} Veljanoski J., et al., 2013b, MNRAS, 2225 

\bibitem[\protect\citeauthoryear{Veljanoski et 
al.}{2014}]{Veljanoskietal14} Veljanoski J., et al., 2014, MNRAS, submitted

\bibitem[\protect\citeauthoryear{Wang 
\& Ma}{2012}]{WangMa12} Wang S., Ma J., 2012, AJ, 143, 132 

\bibitem[\protect\citeauthoryear{Wang 
\& Ma}{2013}]{WangMa13} Wang S., Ma J., 2013, AJ, 146, 20 


\end{thebibliography}
\end{document}